\documentclass[
nofootinbib,
 amsmath,amssymb,
 aps,
floatfix,
]{revtex4-2}

\usepackage{graphicx}
\usepackage{dcolumn}
\usepackage{bm}
\usepackage[utf8]{inputenc}
\usepackage{subfigure}
\usepackage{mathtools}
\usepackage{xcolor}
\usepackage{lineno}

\usepackage{color}

\newcommand{\ket}[1]{\big| #1 \big\rangle}

\newcommand{\braket}[2]{\langle #1 | #2 \rangle}
\newcommand{\vev}[1]{\langle #1 \rangle}

\newcommand{\sn}{\mathrm{sn}}
\newcommand{\cn}{\mathrm{cn}}
\newcommand{\dn}{\mathrm{dn}}

\newcommand{\Sn}{\mathrm{Sn}}
\newcommand{\Cn}{\mathrm{Cn}}
\newcommand{\Dn}{\mathrm{Dn}}

\begin{document}
\title{Constraints on a Generalization of Geometric Quantum Mechanics\\from Neutrino and $B^0$-$\overline{B^0}$ Oscillations}

\author{Nabin Bhatta}
\email{nabinb@vt.edu}

\author{Djordje Minic}
\email{dminic@vt.edu}

\author{Tatsu Takeuchi}
\email{takeuchi@vt.edu}


\affiliation{Center for Neutrino Physics, Department of Physics, Virginia Tech, Blacksburg VA 24061, USA}

\begin{abstract}
Nambu Quantum Mechanics, proposed in Phys. Lett. B536, 305 (2002),
is a deformation of canonical Quantum Mechanics in which the manifold over which the ``phase'' of an energy eigenstate time evolves
is modified.
This generalization affects oscillation and interference phenomena through the introduction of two deformation parameters that quantify the extent of deviation from canonical Quantum Mechanics. In this paper, we constrain these parameters utilizing 
atmospheric neutrino oscillation data, and $B^0$-$\overline{B^0}$ oscillation data from Belle.
Surprisingly, the bound from atmospheric neutrinos is stronger than the bound from Belle.
Various features of Nambu Quantum Mechanics are also discussed.
\end{abstract}


\maketitle

\section{Introduction}
In a previous letter \cite{Minic:2020zjb}, we looked at Nambu Quantum Mechanics proposed in \cite{Minic:2002pd} by Minic and Tze,
which was inspired by an idea of Nambu \cite{Nambu:1973qe}\footnote{Other works motivated by Nambu's paper include \cite{Takhtajan:1993vr,Dito:1996xr,Awata:1999dz,Curtright:2002fd}}, and argued that it could lead to observable consequences
in oscillation phenomena.
In this paper, we provide details of the derivation, expound on the formalism, and place further bounds on the deformation parameters introduced therein.

The objective of this work is to obtain a deeper understanding of canonical Quantum Mechanics (QM).
In particular, we would like to demystify what its foundational principles are.
To this end, we attempt to generalize QM, compare how the predictions of theory will change with respect to those of canonical QM as a consequence of the generalization, 
and confront both with experiment.
This approach will allow us to probe how robust the original tenet or axiom that was relaxed to generalize QM is,
and thereby identify the bedrock on which canonical QM rests. 

Generalization or modification of canonical QM has been attempted in several distinct directions. 
One route has been the replacement of the field over which the state space is defined from complex numbers to other fields or division algebras, i.e. real numbers, quaternions, or octonions, thereby informing us how crucial the complex structure of Hilbert space is to the foundations of QM. 
For a formulation of QM on real vector spaces, see \cite{Stueckelberg:1960}. 
Quaternionic QM is an active field of research and a well-developed formalism can be found in the literature, particularly in the works of Adler \cite{Adler:1986,Adler:1988hb}. 
For a discussion of QM based on the non-associative octonionic algebra see, for instance, \cite{Gunaydin:1978jq}. 
Formulation of QM utilizing vector spaces over finite fields is also possible: See \cite{Chang:2012eh,Chang:2012gg,Chang:2012we,Chang:2013rya,Chang:2019kcp} for 
discrete models of QM over Galois fields. 

Another direction for generalizing canonical QM is relaxing the linearity of the Schr\"odinger equation. Suggestions for testing the linearity of canonical QM exist in the literature, see for instance \cite{Weinberg:1989cm} and \cite{Weinberg:1989us} by Weinberg.  
However, this formulation allows for superluminal communication \cite{Polchinski}. In general, the addition of any non-linear terms to the equations of canonical QM is strongly constrained by experiments. For example, logarithmic additions to the Schr\"odinger equation \cite{Birula} are tightly constrained by neutron interferometry \cite{Zeilinger}.

Yet another research program that generalizes canonical QM is the study of generalized probabilistic theories \cite{Hardy2001quantum,Barrett2007, Janotta_2014,Dakic_2014,PLAVALA20231}. These theories contain canonical QM as a special case and share its non-classical features. In general, this research program includes the study of a broad class of probabilistic theories and the identification of their common operational features. One can then hope to single out the mathematical structure of canonical QM as a special case resulting from some physically motivated axioms, attempted, for instance, by Hardy \cite{Hardy2001quantum}. An example of such a theory that is relevant to our work is Sorkin's hierarchical classification of probabilistic theories according to the order of interference they exhibit \cite{Sorkin:1994dt, Sorkin:1995nj}. In this scheme, both classical probability theory and quantum theory can be seen as special cases of more general probabilistic theories exhibiting interference of orders higher than that of canonical QM. Namely, classical theory is that subset in which there is no interference and canonical QM is characterized by the absence of any interference higher than second order.

Nambu QM generalizes canonical QM via the geometric formulation of QM 
\cite{Kibble:1978tm,Ashtekar:1997ud,Anandan:1990fq,Brody:1999cw,Cirelli:1999in,Cirelli:2003}.
The circular $S^1$ ``phase'' space of the energy eigenstates in geometric QM 
is extended to spherical $S^2$, with the canonical $S^1$ forming the equator of the Nambu $S^2$.
The periodic evolution of the energy eigenstate ``phase'' is deformed from a simple
circular motion along the equator to a more general periodic motion on $S^2$.
This generalization of the ``phase'' space necessarily has consequences on the interference between
energy eigenstates.

The organization of this paper is as follows: 
In Section~\ref{CanonicalQMreview}, we review the basics of the geometric formulation of canonical QM \cite{Kibble:1978tm} and
introduce concepts and notation convenient for describing the Nambu QM extension and contrasting it against canonical QM.
Section~\ref{NambuQMbasics} contains the presentation of Nambu QM as a generalization of geometric QM, 
which can be viewed as a continuous deformation of canonical QM with two deformation parameters.
How this generalization impacts interference and oscillation phenomena is also discussed.
Section~\ref{NeutrinoOscillationBound} compares the formalism developed to atmospheric neutrino data to place bounds
on the deformation parameters, detailing how the results in \cite{Minic:2020zjb} were obtained.
In Section~\ref{BBarBounds}, we contrast how $B^0$-$\overline{B}^0$ oscillations are treated in both canonical and Nambu QM,
and use $B^0$-$\overline{B}^0$ oscillation data from Belle \cite{Belle1} to constrain the deformation parameters of Nambu QM.
The somewhat surprising result is that the bounds on the deformation parameters from atmospheric neutrinos are stronger than those
from Belle.  This is due to the atmospheric neutrino data strongly preferring maximal mixing. 
Section~\ref{Summary} provides a summary of the results, a discussion on the pros and cons of the Nambu approach, and
lists possible future directions of research.
The Appendix includes information that did not fit well into the main text:
In Appendix A we review some properties of Poisson brackets and contrast them with (ternary) Nambu brackets. 
The properties of the asymmetric top are also discussed.
Appendix B lists properties of the Jacobi elliptical functions relevant for this work.

\section{Canonical Quantum Mechanics}\label{CanonicalQMreview}

\subsection{The Geometric Formulation of Canonical Quantum Mechanics}

Due to the rigid and robust structure of canonical QM,
any extension or deformation of its mathematical framework 
necessarily distorts or discards some of its
cherished assumptions and/or principles.
To see which of these properties are compromised or maintained in the
generalization proposed in this paper,
it is worthwhile to start by listing said properties 
in a way that would facilitate the
comparison of canonical QM and the Nambu extension.

\bigskip
\noindent
We maintain the following properties of canonical QM:

\begin{enumerate}
\item
Any quantum state $\ket{\psi}$ of the system under consideration is  described as a superposition of energy eigenstates:
\begin{equation}
\ket{\psi} \;=\; \sum_{n} \psi_n\ket{n}\;.
\end{equation}
Here, we assume for the sake of simplicity that the energy eigenstates are discrete and can be labelled by an integer $n$.
In the following, we will always work in this basis and no other.
We do not consider any change of basis.

\item
In canonical QM, each coefficient $\psi_n$ in the above expansion is an element of the complex
number field $\mathbb{C}$, which can be written as
$\psi_n = A_n e^{i\theta_n}$ where $A_n \in \mathbb{R}$ is the amplitude and $\theta_n\in \mathbb{R}$ is the phase.
Since $e^{i\theta}$ is periodic in $\theta$ with period $2\pi$,
the phase can be restricted to
$\theta\in \mathbb{R}/(2\pi\mathbb{Z}) = [0,2\pi)$.
We do not restrict the amplitude to non-negative reals for latter convenience.

We will generalize this to 
``numbers'' which will maintain the property that
they can be written as
\begin{equation}
\psi_n \;=\; A_n\,x(\theta_n)\;,
\end{equation}
where the $x$-ponent $x(\theta)$ is 
a map from $S^1$ to another manifold which is
periodic in $\theta$ with period $2\pi$:
\begin{equation}
x(\theta+2\pi) \;=\; x(\theta)\;.
\end{equation}
The product of two such ``numbers'' is defined to be:
\begin{equation}
Ax(\alpha) \circ Bx(\beta) \;=\; 
(AB)\;x(\alpha+\beta)\;,
\end{equation}
where $AB$ is just the usual product
between real numbers $A$ and $B$.
It is clear that the $x$-ponent $x(\theta)$
is a representation of $U(1)\cong  SO(2)$. 
However, no sum of the ``numbers'' will be defined, \textit{i.e.}
the ``numbers'' will not comprise a field or
division algebra.
This means that our state space will not retain the full
vector space structure we have in canonical QM.
However, it turns out that this will not be a problem if we work solely in the fixed energy eigenstate basis.

\item The phase $\theta_n$ in the coefficient $\psi_n$ of the $n$-th energy eigenstate time-evolves as
\begin{equation}
\theta_n(t) \;=\; -\omega_n(t-t_n)\;,
\end{equation}
where $\hbar\omega_n$ is identified as the energy of the energy eigenstate $\ket{n}$,
and $t_n$ is the time at which $\theta_n=0$.
We characterize energy eigenstates in this fashion without introducing a Hamiltonian or
Schr\"odinger equation.

When considering unstable states, 
we allow the amplitude $A_n$ to be time dependent as
\begin{equation}
A_n(t) \;=\; A_n(0)\,e^{-(\Gamma_n/2)t}\;.
\end{equation}
We will not add an imaginary part to the energy eigenvalue, since it will not lead
to exponential decay of the coefficient in the Nambu extension.

\item There exists an inner product $\braket{\phi}{\psi}$ between states $\ket{\phi}$ and $\ket{\psi}$ such that
the energy eigenstates are orthonormal, $\braket{n}{m}=\delta_{nm}$,
and
\begin{equation}
\braket{\psi}{\psi}
\;=\; \sum_{n}A_n^2 \;.
\end{equation}
We assume that all stable states are normalized, \textit{i.e.} $\braket{\psi}{\psi}=1$ for all $\ket{\psi}$.
\footnote{
Since the state space of our generalized QM lacks a full vector space structure, referring to $\braket{\phi}{\psi}$ as an ``inner product'' is an abuse of terminology.  However, we will continue to refer to this product as such due to the lack of a better term.
}

\item The Born rule:
When the system is in state $\ket{\psi}$,
the probability of measuring it to be in state $\ket{\phi}$ is given by
\begin{equation}
|\braket{\phi}{\psi}|^2\;.
\label{Born}
\end{equation}

\end{enumerate}


\noindent
We also describe the $x$-ponent and inner product
of canonical QM as follows:
\begin{enumerate}
\setcounter{enumi}{5}
\item 
In canonical QM, the $x$-ponent $x(\theta)$ is $e^{i\theta}$.
This can also be represented as a point
on a unit circle, which we denote:
\begin{equation}
x(\theta) \quad\to\quad 
\vec{x}(\theta) =
\begin{bmatrix}
\cos\theta \\ \sin\theta    
\end{bmatrix}
.
\end{equation}
The coefficient $\psi_n$ is represented by a point on a circle of radius $A_n$:
\begin{equation}
\psi_n \quad\to\quad
\vec{\psi}_n \;=\; 
A_n\,\vec{x}(\theta_n) \;=\;
A_n
\begin{bmatrix}
\cos\theta_n \\ \sin\theta_n    
\end{bmatrix}
.
\label{psivecdef}
\end{equation}
Thus, with an abuse of notation we can write
\begin{equation}
\ket{\psi}\;=\;\sum_{n}\vec{\psi}_n\ket{n}\;.
\end{equation}
Each $\vec{\psi}_n$ will time-evolve clockwise on a circle of
radius $A_n$ with constant angular velocity $\omega_n$.

\item The inner product between two states
\begin{equation}
\ket{\phi} \;=\; \sum_{m}\vec{\phi}_m\ket{m}
\;,
\qquad
\ket{\psi} \;=\; \sum_{n}\vec{\psi}_n\ket{n}
\;,
\end{equation}
is defined as 
\begin{equation}
\braket{\phi}{\psi}
\;=\; g(\phi,\psi) + i\varepsilon(\phi,\psi)\;,
\end{equation}
where
\begin{eqnarray}
g(\phi,\psi) & = & \sum_n (\vec{\phi}_n\cdot\vec{\psi}_n)\;,\cr
\varepsilon(\phi,\psi) & = & \sum_n (\vec{\phi}_n\times\vec{\psi}_n)\;.
\end{eqnarray}
It is straightforward to show that this definition 
agrees with the usual one using complex numbers.
Note that
\begin{equation}
\braket{\psi}{\psi}
\;=\; \sum_{n}(\vec{\psi}_n\cdot\vec{\psi}_n)
\;=\; \sum_{n}A_n^2
\;,
\end{equation}
and
\begin{equation}
|\braket{\phi}{\psi}|^2
\;=\; g(\phi,\psi)^2 + \varepsilon(\phi,\psi)^2 \;.
\end{equation}
Let us write
\begin{equation}
(\vec{\phi}_n\cdot\vec{\psi}_n) \;=\; C_n\cos\zeta_n\;,\qquad
(\vec{\phi}_n\times\vec{\psi}_n) \;=\; C_n\sin\zeta_n\;,
\label{Czetadef}
\end{equation}
that is, $C_n$ is the product of the magnitudes of $\vec{\phi}_n$ and $\vec{\psi}_n$, and
$\zeta_n$ is the angle between the two.
Then
\begin{eqnarray}
|\braket{\phi}{\psi}|^2
& = & \sum_n\sum_m C_n C_m \cos(\zeta_n-\zeta_m) 
\;=\;
\sum_n \underbrace{C_n^2}_{\displaystyle P(n)}
+\sum_{m < n}\underbrace{2C_n C_m\cos(\zeta_n-\zeta_m)}_{\displaystyle I_2(n,m)}
\;,
\label{OneAndTwoPathContributions}
\end{eqnarray}
where $P(n)$ represents the probability that the $n$th energy eigenstate
contributes to $|\braket{\phi}{\psi}|^2$, while $I_2(n,m)$ is the pairwise interference
between the $n$th and $m$th eigenstates a la Sorkin \cite{Sorkin:1994dt}.
Note that the interference $I_2(n,m)$ only depends on the difference $\zeta_n-\zeta_m$.

A phase shift of each energy eigenstate 
\begin{equation}
\ket{n}\quad\to\quad
\vec{x}(\alpha_n)\ket{n}
\;,
\end{equation}
keeps the angle $\zeta_n$ defined in Eq.~\eqref{Czetadef} invariant
since it corresponds to the rotation by angle $\alpha_n$ of the $n$th $S^1$ phase manifold.
Therefore, the inner product $\braket{\phi}{\psi}$ is invariant under such phase shifts.
Consequently, a global phase shift of all superpositions of the energy eigenstates by a common angle $\alpha$,
\begin{equation}
\ket{\psi}\quad\to\quad \vec{x}(\alpha)\ket{\psi}\;,
\end{equation}
leaves the inner product between states invariant.
This property will only be partially maintained in our extension.

In the case of canonical QM,
one can have different phase shifts on each state
\begin{equation}
\ket{\phi}\to \vec{x}(\alpha)\ket{\phi}\;,\qquad
\ket{\psi}\to \vec{x}(\beta)\ket{\psi}\;,
\end{equation}
in which case the symmetric and antisymmetric parts of the 
inner product transform as
\begin{eqnarray}
g(\phi,\psi) & \quad\to\quad & 
g(\phi,\psi)\,\cos(\alpha-\beta)
\;-\;\varepsilon(\phi,\psi)\,\sin(\alpha-\beta)\;,
\vphantom{\big|}\cr
\varepsilon(\phi,\psi) & \quad\to\quad & 
g(\phi,\psi)\,\sin(\alpha-\beta)
\;+\;\varepsilon(\phi,\psi)\,\cos(\alpha-\beta)\;,
\vphantom{\big|}
\end{eqnarray}
leading to the invariance of $|\braket{\psi}{\phi}|^2$.
Thus, an arbitrary change of phase does not lead to any physical consequences, so states that only differ by a phase
are identified as representing the same physical state.
This leads to the projective vector space nature of the canonical QM state space.
This property will be lost in our extension.

\end{enumerate}

\bigskip
In the geometric formulation of canonical QM
\cite{Kibble:1978tm,Ashtekar:1997ud,Anandan:1990fq,Brody:1999cw,Cirelli:1999in,Cirelli:2003}, 
the time evolution of the coefficient 
\begin{equation}
\vec{\psi}_n(t) \;=\; A_n
\begin{bmatrix}
q_n(t) \\ p_n(t)
\end{bmatrix}
\end{equation}
of the $n$th energy eigenstate is viewed as the 
classical evolution of a classical harmonic oscillator along the circle $q_n^2+p_n^2=A_n^2$ in phase space.
For a system with $N$ energy eigenstates, the
system is described by the evolution of a point on a 2$N$-dimensional
K\"ahler manifold with the symmetric and antisymmetric parts
of the inner product respectively 
providing the bilinear and symplectic forms which are connected by
\begin{equation}
g(\phi,\psi) \;=\; \varepsilon(\phi,J\psi)\;,
\end{equation}
where the complex structure $J$ acts on the coefficients $\vec{\psi}_n$ of $\ket{\psi}$ as
\begin{equation}
J\vec{\psi}_n 
\;=\;  
-i\sigma_2 \vec{\psi}_n
\;=\;
\begin{bmatrix}
0 & -1 \\ 1 & \phantom{-}0
\end{bmatrix}\vec{\psi}_n
\;.
\end{equation}
Note that $J^2 = -1$.
Our Nambu extension can be viewed as a generalization
of canonical QM via this geometric formulation.

\subsection{Oscillation}
Before we describe our generalized QM, let us first demonstrate that
the above description of canonical QM allows us to derive the usual
neutrino oscillation formulae.

Let $\ket{\alpha}$ and $\ket{\beta}$ be the flavor eigenstates,
which are defined as superpositions of the
energy eigenstates $\ket{1}$ and $\ket{2}$ via
%
%
\begin{eqnarray}
\ket{\alpha} & = & \phantom{+}c_\theta\ket{1}+s_\theta\ket{2} \;, \cr   
\ket{\beta}  & = & -s_\theta\ket{1}+c_\theta\ket{2} \;,
\end{eqnarray}
where $c_\theta = \cos\theta$, $s_\theta = \sin\theta$.
In our phase vector notation, we have
\begin{equation}
\vec{\alpha}_1 \;=\; c_\theta\,\vec{x}(0)\;,\qquad
\vec{\alpha}_2 \;=\; s_\theta\,\vec{x}(0)\;,\qquad
\vec{\beta}_1 \;=\; -s_\theta\,\vec{x}(0)\;,\qquad
\vec{\beta}_2 \;=\;  c_\theta\,\vec{x}(0)\;,
\end{equation}
where 
\begin{equation}
\vec{x}(0) \;=\;
\begin{bmatrix} 1 \\ 0 \end{bmatrix}
\end{equation}
represents a phaseless state.
Let the state $\ket{\psi}$ at time $t=0$ be equal to $\ket{\alpha}$:
\begin{equation}
\ket{\psi(0)} \;=\; \ket{\alpha} \;=\; c_\theta\ket{1} + s_\theta\ket{2}\;,
\end{equation}
that is
\begin{equation}
\vec{\psi}_1(0) \;=\; \vec{\alpha}_1 \;=\; c_\theta\,\vec{x}(0)\;,\qquad
\vec{\psi}_2(0) \;=\; \vec{\alpha}_2 \;=\; s_\theta\,\vec{x}(0)\;.
\end{equation}
At a later time, these coefficients will have evolved into
\begin{equation}
\vec{\psi}_1(t)
\;=\; c_\theta\,\vec{x}(-\omega_1 t)
\;=\; c_\theta \begin{bmatrix} \cos\omega_1 t \\ -\sin\omega_1 t \end{bmatrix}
\;,\qquad
\vec{\psi}_2(t)
\;=\; s_\theta\,\vec{x}(-\omega_2 t)
\;=\; s_\theta \begin{bmatrix} \cos\omega_2 t \\ -\sin\omega_2 t \end{bmatrix}\;.
\end{equation}
We find
\begin{equation}
\begin{array}{llll}
g(\alpha,\psi(t)) 
& =\;
 \vec{\alpha}_1\cdot\vec{\psi}_1(t)
\!& +\;\vec{\alpha}_2\cdot\vec{\psi}_2(t)
& =\; c_\theta^2\cos\omega_1 t + s_\theta^2\cos\omega_2 t \;,
\vphantom{\Big|}\cr
g(\beta,\psi(t)) 
& =\;
 \vec{\beta}_1\cdot\vec{\psi}_1(t)
\!& +\;\vec{\beta}_2\cdot\vec{\psi}_2(t)
& =\; -s_\theta c_\theta\cos\omega_1 t + s_\theta c_\theta\cos\omega_2 t \;,
\vphantom{\Big|}\cr
\varepsilon(\alpha,\psi(t)) 
& =\; 
 \vec{\alpha}_1\times\vec{\psi}_1(t)
\!& +\;\vec{\alpha}_2\times\vec{\psi}_2(t)
& =\; -c_\theta^2\sin\omega_1 t - s_\theta^2\sin\omega_2 t \;,
\vphantom{\Big|}\cr
\varepsilon(\beta,\psi(t)) 
& =\; 
 \vec{\beta}_1\times\vec{\psi}_1(t)
\!& +\;\vec{\beta}_2\times\vec{\psi}_2(t)
& =\; s_\theta c_\theta\sin\omega_1 t - s_\theta c_\theta\sin\omega_2 t \;.
\vphantom{\Big|}\cr
\end{array}
\end{equation}
Therefore, the survival and transition probabilities will be
\begin{eqnarray}
P(\alpha\to\alpha)
& = & g(\alpha,\psi(t))^2 + \varepsilon(\alpha,\psi(t))^2 
\vphantom{\Big|}\cr
& = &
 \left( c_\theta^2\cos\omega_1 t + s_\theta^2\cos\omega_2 t\right)^2
+\left(-c_\theta^2\sin\omega_1 t - s_\theta^2\sin\omega_2 t\right)^2
\vphantom{\Big|}\cr
& = &
1 - \sin^2 2\theta\, \sin^2\!\left[\dfrac{(\omega_1-\omega_2)t}{2}\right]\;,
\vphantom{\Big|}\cr
P(\alpha\to\beta)
& = & g(\beta,\psi(t))^2 + \varepsilon(\beta,\psi(t))^2 
\vphantom{\Big|}\cr
& = &
 \left(-s_\theta c_\theta\cos\omega_1 t + s_\theta c_\theta\cos\omega_2 t\right)^2
+\left( s_\theta c_\theta\sin\omega_1 t - s_\theta c_\theta\sin\omega_2 t\right)^2
\vphantom{\Big|}\cr
& = & \sin^2 2\theta\, \sin^2\!\left[\dfrac{(\omega_1-\omega_2)t}{2}\right]\;.
\vphantom{\Big|}
\end{eqnarray}
These are the usual oscillation equations.
To render the expression relativistic, we make the replacement
\begin{equation}
\omega_i t 
\quad\to\quad (\omega_i t - k_i L) 
\quad\xrightarrow{\mathrm{natural\ units}}\quad
(E_i t - p_i L)\;.
\end{equation}
If the energies are common, $E_1=E_2=E$, but $m_1\neq m_2$, then
\begin{eqnarray}
(E t - p_i L)
& = & E t - \sqrt{E^2 -m_i^2}\,L
\cr
& \approx & E t - E\left(1 - \dfrac{m_i^2}{2E^2}\right) L 
\;=\; E(t-L) + \dfrac{m_i^2}{2E}L\;.
\end{eqnarray}
Therefore, the replacement is
\begin{equation}
(\omega_1-\omega_2)t
\quad\to\quad \dfrac{\delta m_{12}^2}{2E}\,L
\;\equiv\; \Delta \;,
\label{DeltaDef}
\end{equation}
which precisely recovers the usual neutrino oscillation relations \cite{Gonzalez-Garcia:2020}.

\bigskip
\section{Nambu Quantum Mechanics}\label{NambuQMbasics}

\begin{figure}[ht]
\includegraphics[width=7cm]{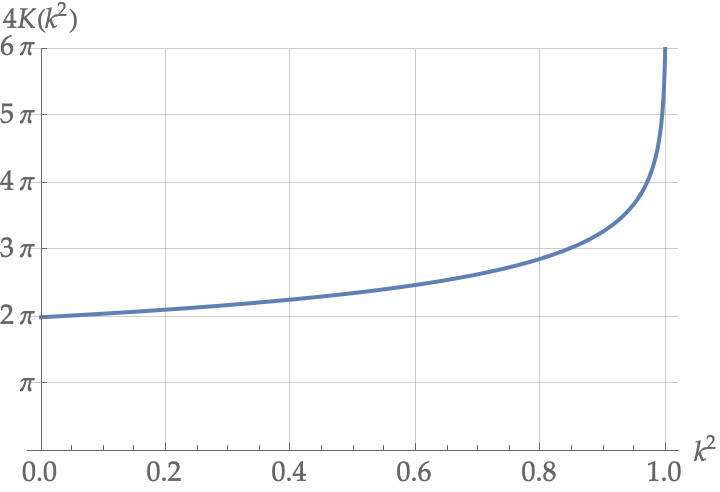}
\caption{
The $k^2$-dependence of $K=K(k^2)$,
the complete elliptical integral of the first kind.
Shown is the graph for $4K$, which is the period of $\sn(u,k)$ and $\cn(u,k)$
in $u$.
Note that $K\to\infty$ as $k^2\to 1$.
}
\label{Km}
\end{figure}

\begin{figure}[ht]
\includegraphics[width=17.5cm]{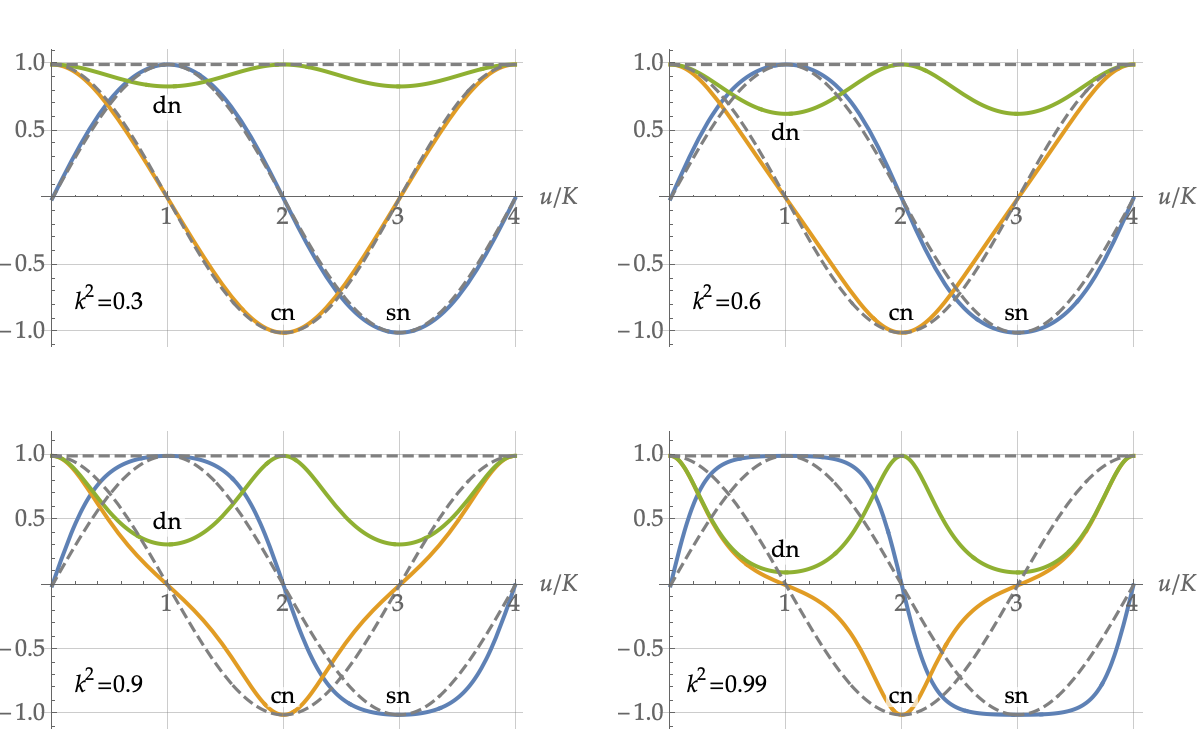}
\caption{
The $u$-dependence of the Jacobi elliptical functions
$\sn(u,k)$, $\cn(u,k)$, and $\dn(u,k)$
for $k^2=0.3$ (upper left), $k^2=0.6$ (upper right), $k^2=0.9$ (lower left), and $k^2=0.99$ (lower right).
The horizontal axis is in units of $K=K(k^2)$, 
the complete elliptical integral of the first kind.
The dashed lines are the $k^2=0$ case where
$\sn(u,0)=\sin u$, $\cn(u,0)=\cos u$, $\dn(u,0)=1$.
}
\label{sncndn}
\end{figure}

\subsection{Deformation of Quantum Mechanics into a theory of Asymmetric Tops}

The deformation of canonical QM discussed in Refs.~\cite{Minic:2002pd,Minic:2020zjb} can be summarized as
follows:
The $x$-ponent, which is a map
from $S^1$ to $S^1$ in canonical QM,
is deformed to a map from $S^1$ to
$S^2$ as:
\begin{equation}
\vec{x}(\theta)
=
\begin{bmatrix}
\cos\theta \\ \sin\theta   
\end{bmatrix}
\qquad\to\qquad
\vec{X}(\theta|k,\xi) = 
\begin{bmatrix}
c_\xi\; \cn(2K\theta/\pi,k) \vphantom{\Big|}\\
\sqrt{c_\xi^2 + s_\xi^2 k^2}\;\sn(2K\theta/\pi,k) \\
-s_\xi\; \dn(2K\theta/\pi,k) \vphantom{\Big|}
\end{bmatrix}
\;,
\label{U1map2S2}
\end{equation}
where $\cn(u,k)$, $\sn(u,k)$, $\dn(u,k)$, with $0\le k < 1$, are Jacobi's elliptical functions \cite{Abramowitz-Stegun}\footnote{
In the literature, Jacobi's elliptical functions
are also written as $\cn(u,m)$, $\sn(u,m)$, and $\dn(u,m)$ with $m=k^2$, so care is necessary when comparing formulae.
The functions JacobiCN[u,m], JacobiSN[u,m], JacobiDN[u,m] on Mathematica also adopt this definition.
}
, and 
$c_\xi=\cos\xi$, $s_\xi=\sin\xi$,  where 
$0 \le \xi\le \pi/2$ is the second
deformation parameter.
The reason for the minus sign on the third component is explained later.
Note that Jacobi's elliptical functions satisfy the relations
\begin{eqnarray}
\cn^2(u,k) + \sn^2(u,k) & = & 1 \;,\vphantom{\big|}\cr
k^2\sn^2(u,k) + \dn^2(u,k) & = & 1\;,\vphantom{\big|}
\end{eqnarray}
so $\vec{X}(\theta|k,\xi)$ has unit norm.

The periods of $\cn(u,k)$ and $\sn(u,k)$ for real $u$ is $4K$ while that of $\dn(u,k)$ is $2K$,
where $K=K(k^2)$ is the complete elliptical integral of the first kind \cite{Abramowitz-Stegun}. 
See FIGS.~\ref{Km} and \ref{sncndn}.
In the above map, the arguments of the elliptical
functions are rescaled so that the period in $\theta$ is
$2\pi$ for $\cn$ and $\sn$, and $\pi$ for $\dn$.
For latter convenience, we introduce the notation 
\begin{equation}
\Cn(\theta,k) \,\equiv\, \cn(2K\theta/\pi,k)\;,\qquad
\Sn(\theta,k) \,\equiv\, \sn(2K\theta/\pi,k)\;,\qquad
\Dn(\theta,k) \,\equiv\, \dn(2K\theta/\pi,k)\;,
\end{equation}
to refer to these functions with their arguments rescaled.
Due to the period of $\Dn(\theta)$ being $\pi$ instead of $2\pi$, 
a phase shift by $\pi$ does not change the sign of $\vec{X}(\theta|k,\xi)$:
\begin{equation}
\vec{X}(\theta+\pi|k,\xi) \;\neq\; -\vec{X}(\theta|k,\xi)\;,
\end{equation}
except when $\xi=0$ in which case the third component of $\vec{X}(\theta|k,0)$ is missing.
To allow for a sign flip, we let the magnitudes of the coefficients be negative as well as positive.

\begin{figure}[ht]
\begin{center}
\includegraphics[width=10cm]{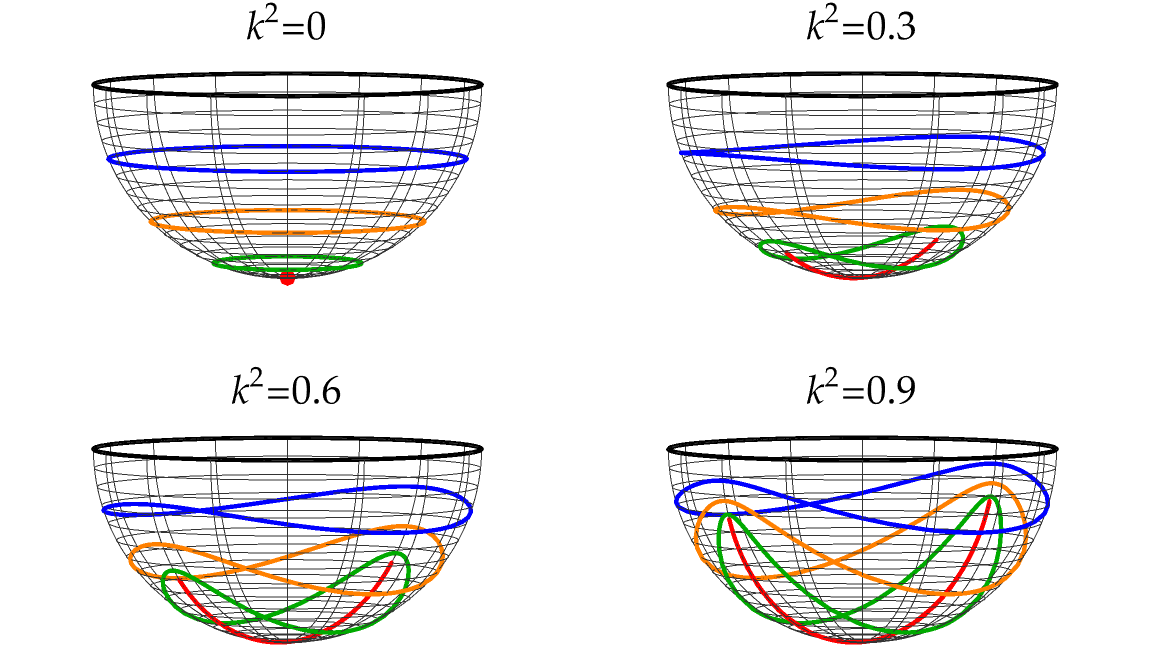}
\end{center}
\caption{Phase trajectories on $S^2$ in Nambu QM for positive amplitudes. Only the southern hemisphere is shown.
Trajectories in the northern hemisphere occur when the amplitude is negative.
The colored lines in each figure indicate $\xi=0$ (Black), $\xi=\pi/8$ (Blue), $\xi=\pi/4$ (Orange), $\xi=3\pi/8$ (Green), and
$\xi=\pi/2$ (Red).  
When $\xi=0$, the trajectory always follows the equator regardless of the value of $k$.
However, the phase time-evolves with uniform angular velocity along the equator only when $k=0$.
When $k=0$ and $\xi=\pi/2$, the phase is stationary at the south pole.
}
\label{PhaseTrajectories}
\end{figure}

The dependence of the trajectory of $\vec{X}(\theta|k,\xi)$ on $S^2$ on the deformation parameters $k$ and $\xi$ is shown FIG.~\ref{PhaseTrajectories}.
For instance, when $k=0$ the trajectory reduces to
\begin{equation}
\vec{X}(\theta|0,\xi)\;=\;
\begin{bmatrix}
c_{\xi}\vec{x}(\theta) \\ -s_{\xi}    
\end{bmatrix}
\;=\;
\begin{bmatrix}
c_\xi\cos\theta \\ c_\xi\sin\theta \\ -s_\xi    
\end{bmatrix}
\;,
\end{equation}
namely, the path of constant latitude $-\xi$ on $S^2$.
When $\xi=0$ this will be the equator, whereas when $\xi=\pi/2$ this will be the south pole.
Setting $\xi=0$ for a non-zero $k$ would lead to
\begin{equation}
\vec{X}(\theta|k,0)
\;=\;
\begin{bmatrix}
\Cn(\theta,k) \\ \Sn(\theta,k) \\ 0    
\end{bmatrix}
\;,
\end{equation}
which traces the equator of $S^2$ but with a non-canonical dependence on $\theta$.

The coefficients of the $n$th energy eigenstate will time-evolve as
\begin{equation}
\vec{\Psi}_n(t) 
\;=\;
A_n\,\vec{X}(-\omega_n(t-t_n)|k_n,\xi_n)
\;=\;
A_n
\begin{bmatrix} 
\phantom{-}c_{\xi_n}\Cn(\omega_n(t-t_n),k_n) \vphantom{\Big|}\\ 
         -\sqrt{c_{\xi_n}^2+s_{\xi_n}^2 k_n^2}\;\Sn(\omega_n(t-t_n),k_n) \\ 
         -s_{\xi_n}\Dn(\omega_n(t-t_n),k_n) 
\vphantom{\Big|}
\end{bmatrix},
\label{NambuTimeEvolution}
\end{equation} 
where $t_n$ sets the initial condition for each $n$.
Note that when $\xi=0$ the trajectory of 
$\vec{X}(-\omega t|k,0)$ will be along the equator of $S^2$, but it will not time-evolve
with constant angular velocity $\omega$ except for the canonical QM case $k=0$.

The trajectory of $\vec{\Psi}_n(t)$ on $S^2$ is that of the angular momentum of a free asymmetric top in the frame fixed to itself.
There, the angular momentum
\begin{equation}
A^2 \;=\; L_1^2 + L_2^2 + L_3^2\;,
\end{equation}
and energy
\begin{equation}
E \;=\;
  \dfrac{L_1^2}{2I_1} 
+ \dfrac{L_2^2}{2I_2} 
+ \dfrac{L_3^2}{2I_3}
\;,
\end{equation}
are both fixed so the angular momentum vector will evolve along the intersection of the two surfaces
$A^2=$constant and $E=$constant.  $k=0$ corresponds to the symmetric top with $I_1=I_2$.
However, the dynamics we chose in Eq.~\eqref{NambuTimeEvolution} is not that of the asymmetric top.  
Indeed, when the intersection of the two surfaces is the equator, $\xi=0$, the angular 
momentum of the asymmetric (symmetric at the equator) top stays fixed and does not time evolve.
This must be the case since the $\xi=0$ limit must correspond to the same physical configuration for the top regardless of the value of $k$.
See Appendix~\ref{Asymtop} for details.
Our dynamics here is chosen so that $k=0$ and $\xi=0$ recovers canonical QM.
The only thing that is maintained is the direction of evolution along the trajectory, by the choice of the
sign of the third component.

Note that this is but one way to generate a closed
path on $S^2$. 
Other deformations of the $x$-ponent are possible,
each leading to different generalizations of canonical QM.  
However, we will not dwell on this question further in this paper.

\subsection{Inner Product}

We extend the inner product of two states to
the quaternion
\begin{equation}
\braket{\Phi}{\Psi}
\;=\; g(\Phi,\Psi)
- i\,\varepsilon_1(\Phi,\Psi)
- j\,\varepsilon_2(\Phi,\Psi)
- k\,\varepsilon_3(\Phi,\Psi)
\;,
\end{equation}
where
\begin{eqnarray}
g(\Phi,\Psi) & = & \sum_{n}(\vec{\Phi}_n\cdot\vec{\Psi}_n)\;,
\vphantom{\bigg|}\cr
\vec{\varepsilon}\,(\Phi,\Psi) 
& = &
\begin{bmatrix}
\varepsilon_1(\Phi,\Psi) \\
\varepsilon_2(\Phi,\Psi) \\
\varepsilon_3(\Phi,\Psi)
\end{bmatrix}
\;=\; \sum_{n}(\vec{\Phi}_n\times\vec{\Psi}_n)\;.
\vphantom{\bigg|}
\end{eqnarray}
The dot and cross products are now defined in three dimensions.
This inner product results if we map the 3D phase vector $\vec{\Psi}$ to a purely imaginary quaternion
\begin{equation}
\vec{\Psi}\,=\,
\begin{bmatrix}
\Psi_1 \\ \Psi_2 \\ \Psi_3
\end{bmatrix}
\qquad\to\qquad
\Psi\,=\, i\Psi_1 + j\Psi_2 + k\Psi_3\;,
\end{equation}
and define the inner product to be
\begin{eqnarray}
\braket{\Phi}{\Psi}
& = & \overline{\Phi}\Psi
\;=\;
-(i\Phi_1 + j\Phi_2 + k\Phi_3)(i\Psi_1 + j\Psi_2 + k\Psi_3)
\;.
\end{eqnarray}
Note that
\begin{equation}
\braket{\Psi}{\Psi}
\;=\; g(\Psi,\Psi)
\;=\; \sum_{n}A_n^2 
\;,
\end{equation}
and
\begin{equation}
|\braket{\Phi}{\Psi}|^2
\;=\; g(\Phi,\Psi)^2
+ 
\vec{\varepsilon}\,(\Phi,\Psi)\cdot
\vec{\varepsilon}\,(\Phi,\Psi)
\;.
\label{braketPhiPsiSquared}
\end{equation}
If we write
\begin{equation}
(\vec{\Phi}_n\cdot\vec{\Psi}_n)\;=\;C_n\cos\zeta_n\;,\qquad
(\vec{\Phi}_n\times\vec{\Psi}_n)\;=\;\vec{C}_n\sin\zeta_n\;,
\label{CzetaNambudef}
\end{equation}
where $|\vec{C}_n|=C_n$, then
\begin{eqnarray}
|\braket{\Phi}{\Psi}|^2
& = & \sum_n\sum_m 
\Big(C_n C_m \cos\zeta_n\cos\zeta_m
+\vec{C}_n\cdot\vec{C}_m\sin\zeta_n\sin\zeta_m
\Big)
\cr
& = & \sum_n \underbrace{C_n^2}_{\displaystyle P(n)}
+ \sum_{n<m}
\underbrace{2C_n C_m\Big(\cos\zeta_n\cos\zeta_m + \cos\gamma_{nm}\sin\zeta_n\sin\zeta_m\Big)}_{\displaystyle I_2(n,m)}
\;,
\end{eqnarray}
where $\gamma_{nm}$ is the angle between $\vec{C}_n$ and $\vec{C}_m$.
Unlike the canonical QM case, \textit{cf.} Eq.~\eqref{OneAndTwoPathContributions},
the interference $I_2(n,m)$ will depend on the sum $\zeta_n+\zeta_m$ as well as the difference $\zeta_n-\zeta_m$
unless $\cos\gamma_{nm}=1$.  

The angle $\zeta_n$ defined in Eq.~\eqref{CzetaNambudef} is invariant under 3D rotations of the $n$th
$S^2$ phase manifold associated with the $n$th energy eigenstate,
generalizing the canonical case which was invariant under 2D rotations of the $S^1$ phase manifold.
Note, however, that in general, a phase shift of an energy eigenstate
\begin{equation}
\ket{n} \quad\to\quad
\vec{X}(\alpha_n|k_n,\xi_n)\ket{n}
\end{equation}
does not correspond to a 3D rotation of $S^2$, except for the symmetric top case $k_n=0$, in which case this phase shift becomes a rotation of
angle $\alpha$ around the north-south axis of $S^2$.
Thus, to maintain the invariance of the angles $\zeta_n$, and consequently that of the inner product $\braket{\Phi}{\Psi}$ 
under phase shifts of the individual energy eigenstates, we must choose $k_n=0$ for all $n$.
Even then, we do not have the freedom to shift the phases of different states, \textit{i.e.} different linear combinations of the
energy eigenstates, independently since $|\braket{\Phi}{\Psi}|^2$ will not remain invariant.
So the projective nature of the state space is lost.

This difference between the inner products in canonical QM and Nambu QM has consequences in oscillation phenomena. 
We demonstrate this using the calculation of 
neutrino oscillation probabilities next, allowing
$k_n\neq 0$ to see their effect.

\subsection{Neutrino Oscillations in Nambu QM}

As an example of oscillation phenomena, 
we calculate the two-flavor neutrino oscillation probabilities. 
For simplicity, we take the deformation parameters $k_n$ and $\xi_n$ to be the same for all energy eigenstates.

Let the flavor eigenstates $\ket{A}$ and $\ket{B}$
be given by superpositions of mass eigenstates
$\ket{1}$ and $\ket{2}$ as
\begin{eqnarray}
\ket{A} & = & 
\phantom{+}c_\theta \ket{1} + s_\theta \ket{2}\;,
\vphantom{\Big|}\cr
\ket{B} & = & 
-s_\theta \ket{1} + c_\theta \ket{2}\;.
\vphantom{\Big|}
\label{FlavorEigenstates}
\end{eqnarray}
Then,
\begin{equation}
\vec{A}_1 \;=\;  c_\theta\,\vec{X}(0)\;,\qquad
\vec{A}_2 \;=\;  s_\theta\,\vec{X}(0)\;,\qquad
\vec{B}_1 \;=\; -s_\theta\,\vec{X}(0)\;,\qquad
\vec{B}_2 \;=\;  c_\theta\,\vec{X}(0)\;,
\end{equation}
where
\begin{equation}
\vec{X}(0)
\;=\;
\vec{X}(0|k,\xi)
\;=\;
\begin{bmatrix} c_\xi \\ 0 \\ -s_\xi \end{bmatrix}
\end{equation}
represents a ``phaseless'' state.
Let $\ket{\Psi(0)} = \ket{A}$, that is:
\begin{equation}
\vec{\Psi}_1(0) \;=\; \vec{A}_1 \;=\; c_\theta\,\,\vec{X}(0)\;,\qquad
\vec{\Psi}_2(0) \;=\; \vec{A}_2 \;=\; s_\theta\,\,\vec{X}(0)\;.
\end{equation}
At a later time $t$, we have
\begin{equation}
\vec{\Psi}_1(t) \;=\; c_\theta\,\,
\vec{X}(-\omega_1 t)\;,\qquad
\vec{\Psi}_2(t) \;=\; s_\theta\,
\vec{X}(-\omega_2 t)\;,
\end{equation}
where
\begin{equation}
\vec{X}(-\omega_i t) \;=\;
\vec{X}(-\omega_i t|k,\xi) \;=\;
\begin{bmatrix}
c_{\xi}\, \Cn(\omega_i t,k) \\
-\kappa c_{\xi}\, \Sn(\omega_i t,k) \\
-s_{\xi}\, \Dn(\omega_i t,k)
\end{bmatrix}
\;=\;
\begin{bmatrix}
c_{\xi}\, \Cn_i \\
-\kappa c_{\xi}\, \Sn_i \\
-s_{\xi}\, \Dn_i
\end{bmatrix}\;.
\end{equation}
Here, we use the shorthand
\begin{equation}
\kappa \,=\, \sqrt{1+k^2 \tan^2\xi}\;,\quad
\Sn_i \,=\, \Sn(\omega_i t,k)\;,\quad
\Cn_i \,=\, \Cn(\omega_i t,k)\;,\quad
\Dn_i \,=\, \Dn(\omega_i t,k)\;.
\end{equation}
Then, the symmetric parts of $\braket{A}{\Psi(t)}$ and $\braket{B}{\Psi(t)}$ are
\begin{eqnarray}
g(A,\Psi(t))
& = & \vec{A}_1\cdot\vec{\Psi}_1(t) + \vec{A}_2\cdot\vec{\Psi}_2(t)
\vphantom{\bigg|}\cr
& = & 
  c_\theta^2\,\vec{X}(0)\cdot\vec{X}(-\omega_1 t) 
+ s_\theta^2\,\vec{X}(0)\cdot\vec{X}(-\omega_2 t)
\vphantom{\bigg|}\cr
& = & c_\theta^2\Bigl(\,
c_\xi^2\, \Cn_1 + s_\xi^2\, \Dn_1
\,\Bigr)
+ s_\theta^2\Bigl(\,
c_\xi^2\, \Cn_2 + s_\xi^2\, \Dn_2
\,\Bigr)
\;,\vphantom{\bigg|}\cr
g(B,\Psi(t))
& = & \vec{B}_1\cdot\vec{\Psi}_1(t) + \vec{B}_2\cdot\vec{\Psi}_2(t)
\vphantom{\bigg|}\cr
& = & -s_\theta c_\theta\,
\vec{X}(0)\cdot\vec{X}(-\omega_1 t) 
+ s_\theta c_\theta\,
\vec{X}(0)\cdot\vec{X}(-\omega_2 t)
\vphantom{\bigg|}\cr
& = & -s_\theta c_\theta\Bigl(\,
c_\xi^2\, \Cn_1 + s_\xi^2\, \Dn_1
\,\Bigr)
+ s_\theta c_\theta\Bigl(\,
c_\xi^2\, \Cn_2 + s_\xi^2\, \Dn_2
\,\Bigr)
\;.\vphantom{\bigg|}
\end{eqnarray}
while the antisymmetric parts of $\braket{A}{\Psi(t)}$ and $\braket{B}{\Psi(t)}$ are
\begin{eqnarray}
\vec{\varepsilon}\,(A,\Psi(t))
& = & \vec{A}_1\times\vec{\Psi}_1(t) + \vec{A}_2\times\vec{\Psi}_2(t)
\cr
& = & c_\theta^2\,
\vec{X}(0)\times\vec{X}(-\omega_1 t) 
+ s_\theta^2\,
\vec{X}(0)\times\vec{X}(-\omega_2 t)
\vphantom{\bigg|}\cr
& = & c_\theta^2
\begin{bmatrix}
-\kappa s_\xi c_\xi\,\Sn_1 \\
s_\xi c_\xi \bigl( \Dn_1 - \Cn_1 \bigr) \\
-\kappa c_\xi^2\, \Sn_1
\end{bmatrix}
+ s_\theta^2
\begin{bmatrix}
-\kappa s_\xi c_\xi\,\Sn_2 \\
s_\xi c_\xi \bigl( \Dn_2 - \Cn_2 \bigr) \\
-\kappa c_\xi^2\, \Sn_2
\end{bmatrix}
\;,\cr
& & \vphantom{x}\cr
\vec{\varepsilon}\,(B,\Psi(t))
& = & \vec{B}_1\times\vec{\Psi}_1(t) + \vec{B}_2\times\vec{\Psi}_2(t)
\cr
& = & -s_\theta c_\theta\;
\vec{X}(0)\times\vec{X}(-\omega_1 t) 
+ s_\theta c_\theta\;
\vec{X}(0)\times\vec{X}(-\omega_2 t)
\vphantom{\bigg|}\cr
& = & -s_\theta c_\theta
\begin{bmatrix}
-\kappa s_\xi c_\xi\,\Sn_1 \\
s_\xi c_\xi \bigl( \Dn_1 - \Cn_1 \bigr) \\
-\kappa c_\xi^2\, \Sn_1
\end{bmatrix}
+ s_\theta c_\theta
\begin{bmatrix}
-\kappa s_\xi c_\xi\,\Sn_2 \\
s_\xi c_\xi \bigl( \Dn_2 - \Cn_2 \bigr) \\
-\kappa c_\xi^2\, \Sn_2
\end{bmatrix}
\;.
\label{PAAPAB-Nambu}
\end{eqnarray}
Therefore, the survival and transition probabilities are given by
\begin{eqnarray}
P(A\to A) & = & g(A,\Psi(t))^2 + \vec{\varepsilon}\,(A,\Psi(t))\cdot \vec{\varepsilon}\,(A,\Psi(t)) 
\vphantom{\Big|}\cr
& = & c_\theta^4 + s_\theta^4 
+2c_\theta^2 s_\theta^2
\Bigl[
c_\xi^2
\big( \Sn_1\,\Sn_2 + \Cn_1\,\Cn_2 \bigr)
+ s_\xi^2
\big( k^2\,\Sn_1\,\Sn_2 + \Dn_1\,\Dn_2 \bigr)
\Bigr]
\vphantom{\Big|}\cr
& = & 1 
- \sin^2 2\theta
\left[
\dfrac{1
-\bigl\{
 c_\xi^2
\bigl( \Sn_1\,\Sn_2 + \Cn_1\,\Cn_2 \bigr)
+s_\xi^2
\bigl( k^2\,\Sn_1\,\Sn_2 + \Dn_1\,\Dn_2 \bigr)
\bigr\}
}
{2}
\right]
\;,\cr
& & \phantom{x}\cr
P(A\to B) & = & g(B,\Psi(t))^2 + \vec{\varepsilon}\,(B,\Psi(t))\cdot \vec{\varepsilon}\,(B,\Psi(t)) 
\vphantom{\Big|}\cr
& = &
2c_\theta^2 s_\theta^2
\Bigl[1
-c_\xi^2
\big( \Sn_1\,\Sn_2 + \Cn_1\,\Cn_2 \bigr)
-s_\xi^2
\big( k^2\,\Sn_1\,\Sn_2 + \Dn_1\,\Dn_2 \bigr)
\Bigr]
\vphantom{\Big|}\cr
& = & 
\sin^2 2\theta
\left[
\dfrac{1
-
\bigl\{c_\xi^2
\bigl( \Sn_1\,\Sn_2 + \Cn_1\,\Cn_2 \bigr)
+s_\xi^2
\bigl( k^2\,\Sn_1\,\Sn_2 + \Dn_1\,\Dn_2 \bigr)
\bigr\}
}
{2}
\right]
\;.
\label{PAAandPAB}
\end{eqnarray}
Note that the conservation of probability is manifest:
\begin{equation}
P(A\to A) \,+\, P(A\to B) \;=\; 1\;.
\end{equation}
Let us define
\begin{equation}
\omega \,\equiv\, \dfrac{\omega_1+\omega_2}{2}\;,\qquad
\delta\omega \,\equiv\, \omega_1-\omega_2\;,
\end{equation}
so that we can write
\begin{equation}
\omega_1 \,=\, \omega + \dfrac{\delta\omega}{2}\;,\qquad
\omega_2 \,=\, \omega - \dfrac{\delta\omega}{2}\;.
\end{equation}
Utilizing the addition and subtraction theorems given in Appendix~\ref{Jacobi},
Eq.~\eqref{AdditionSubtraction2}, 
we can rewrite the
combinations $(\Sn_1\,\Sn_2 + \Cn_1\,\Cn_2)$ and $(k^2\,\Sn_1\,\Sn_2 + \Dn_1\,\Dn_2)$
as functions of $\omega t$ and $\delta\omega\,t$,
\begin{eqnarray}
\Sn_1\,\Sn_2 + \Cn_1\,\Cn_2
& = & 
1 -\dfrac{2\,\Dn^2(\omega t)\,\Sn^2(\delta\omega\,t/2)}{1-k^2\,\Sn^2 (\omega t)\,\Sn^2(\delta\omega\,t/2)}
\;,
\vphantom{\Bigg|}\cr
k^2\,\Sn_1\,\Sn_2 + \Dn_1\,\Dn_2
& = &
1 -\dfrac{2k^2\Cn^2(\omega t)\,\Sn^2(\delta\omega\,t/2)}{1-k^2\,\Sn^2 (\omega t)\,\Sn^2(\delta\omega\,t/2)}
\;,
\vphantom{\Bigg|}
\end{eqnarray}
and we find
\begin{eqnarray}
\mathcal{F}(\omega t,\delta\omega\, t)
& \equiv &
\dfrac{1
-
\bigl\{c_\xi^2
\bigl( \Sn_1\,\Sn_2 + \Cn_1\,\Cn_2 \bigr)
+s_\xi^2
\bigl( k^2\,\Sn_1\,\Sn_2 + \Dn_1\,\Dn_2 \bigr)
\bigr\}
}
{2}
\vphantom{\Bigg|}
\cr
& = & 
\dfrac{(c_\xi^2 + s_\xi^2 k^2) - k^2\Sn^2(\omega t)}
{1-k^2\,\Sn^2 (\omega t)\,\Sn^2(\delta\omega\,t/2)}
\;\Sn^2(\delta\omega\,t/2)
\;.
\vphantom{\Bigg|}
\label{Fdef}
\end{eqnarray}
To render this expression
relativistic and appropriate
for neutrino oscillations, we make the replacement
\begin{eqnarray}
\omega_i t 
& \to &
Et - \sqrt{E^2 - m_i^2}\,L
\cr
& \approx &
E(t-L) + \dfrac{m_i^2}{2E}L
\;,
\label{uReplacement}
\end{eqnarray}
where we assume $E\gg m_i$. 
Note that
\begin{eqnarray}
\omega\,t
& \to & 
E(t-L)
\;,
\cr
\delta\omega\,t 
& \to &
\dfrac{\delta m_{12}^2}{2E}L
\;=\; \Delta
\;.
\end{eqnarray}
Thus, for $\mathcal{F}(\omega t,\Delta)$ to depend
only on $\Delta$, we must set $k=0$, 
in which case we have
\begin{equation}
\mathcal{F}(\omega t,\Delta)
\quad\xrightarrow{k\to 0}\quad
c_\xi^2\sin^2\dfrac{\Delta}{2}
\;.
\end{equation}
Other than the factor of
$c_\xi^2$, this is the same as the
expression for canonical QM.
However, due to the $c_\xi^2$ factor, 
oscillations vanish in the limit $\xi\to\frac{\pi}{2}$.

Note that when $k=0$, $\xi=\pi/2$, the phase vectors of all energy eigenstates will be pointing to the
south pole of the $S^2$ phase space, and do not evolve with time.
The angles between arbitrary pairs of phase vectors will always be zero, and consequently,
the asymmetric part of the inner product, $\vec{\varepsilon}(\Phi,\Psi)$, will be identically zero.
It is tempting to think of this as the ``classical'' limit of the theory, in which
all oscillation phenomena vanish.  However, pairwise interferences do not vanish altogether in this limit
since they continue to exist in the $g(\Phi,\Psi)^2$ contribution to $|\braket{\Phi}{\Psi}|^2$.
Indeed, these interferences are necessary to maintain unitarity.
Due to this, we will call the $k=0$, $\xi=\pi/2$ case ``pseudo-classical.''

\begin{figure}[ht]
\begin{center}
\includegraphics[width=17.5cm]{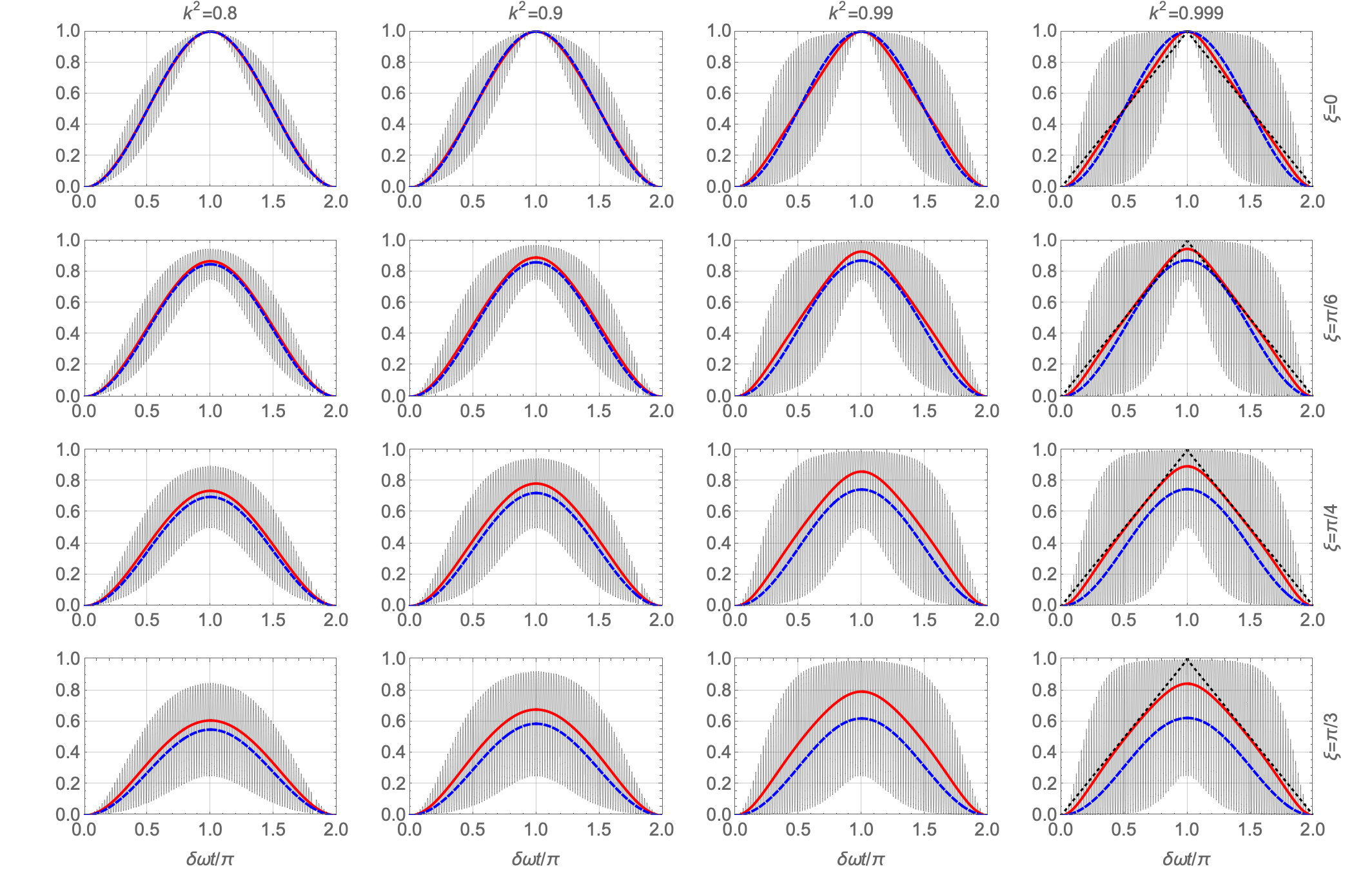}
\end{center}
\caption{The graph of
$\mathcal{F}(\omega t,\delta\omega\,t)$ when
$2\omega/\delta\omega =(\omega_1+\omega_2)/(\omega_1-\omega_2)=100$,
for the cases
$k^2=0.8$, $0.9$, $0.99$, $0.999$ and
$\xi=0$, $\frac{\pi}{6}$,
$\frac{\pi}{4}$, $\frac{\pi}{3}$.
Graphs for the same value of $k^2$ are arranged in columns while those for the same value of $\xi$ are arranged in rows.
Red: graph of $\overline{\mathcal{F}}(\delta\omega\,t)$ which averages
over the rapid oscillations of $\mathcal{F}(\omega t,\delta\omega\,t)$.
Blue dashed: graph of Eq.~\eqref{nosc}.
Note that the red graph is well approximated by the blue, especially for small $\xi$, until $k^2$
closely approaches 1.
In the limit $k^2\to 1$,
the locally averaged graph (red line) converges to the saw-tooth function indicated with the black dotted line on the graphs in the rightmost column ($k^2=0.999$).
}
\label{NQMoscillations}
\end{figure}

For $k\neq 0$ the dependence on $\omega t$ remains.
However, when $E\gg m_i$ we have
$\omega \gg \delta\omega$, 
and in the expression for $\mathcal{F}(\omega t,\delta\omega\,t)$ given in Eq.~\eqref{Fdef}, 
$\Sn^2(\omega t,k)$ will oscillate very rapidly compared to $\Sn^2(\delta\omega\,t/2,k)$.
We expect such rapid oscillations to be unobservable, and that
$\mathcal{F}(\omega t,\Delta)$ can be replaced by the function
$\overline{\mathcal{F}}(\Delta)$ where
\begin{equation}
\overline{\mathcal{F}}(y) 
\;=\; \dfrac{1}{2\pi}\int_{-\pi}^{\pi}dx\,\mathcal{F}(x,y)\;.
\end{equation}
We demonstrate this in FIG.~\ref{NQMoscillations}
which shows graphs of $\mathcal{F}(\omega t,\delta\omega\,t)$
for several choices of $k^2$ and $\xi$
for the case
\begin{equation}
\dfrac{2\omega}{\delta\omega}
\;=\;
\dfrac{\omega_1+\omega_2} 
      {\omega_1-\omega_2}
\;=\; 100\;.
\end{equation}
We can see that
$\mathcal{F}(\omega t,\delta\omega\,t)$ oscillates
rapidly around the more slowly varying function
$\overline{\mathcal{F}}(\delta\omega\,t)$
shown in red. 
Using the expansion of the Jacobi elliptic functions in $k^2$ given in Appendix~\ref{Jacobi},  
we can show that $\overline{\mathcal{F}}(\delta\omega\,t)$ can be approximated by
\begin{eqnarray}
\overline{\mathcal{F}}(\delta\omega\,t)
& = &
\bigg(
c_\xi^2 + s_\xi^2 \dfrac{k^2}{2}
\bigg)
\sin^2\dfrac{\Delta}{2}
+ O(k^4)\;.
\label{nosc}
\end{eqnarray}
This graph is also shown on FIG.~\ref{NQMoscillations} in
dashed blue, 
and we see that this approximation breaks down only when $k^2$ is very close to 1.

\section{Bound from Atmospheric Neutrinos}\label{NeutrinoOscillationBound}

\begin{figure}[ht]
\includegraphics[width=7cm]{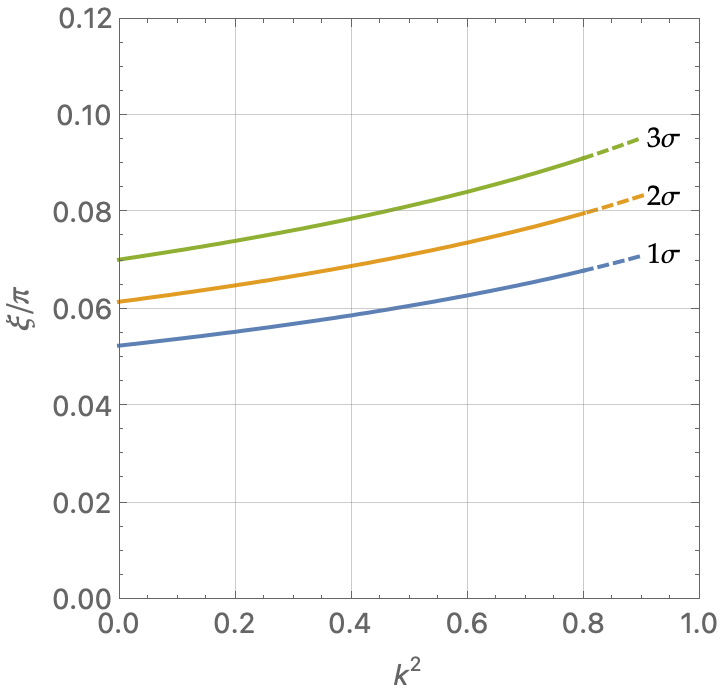}
\caption{$k^2$-$\xi$ contour plot showing the $1\sigma$ (blue), 
$2\sigma$ (orange), and $3\sigma$ (green) bounds
from atmospheric neutrino data
and Eq.~\eqref{nosc}, 
assuming Normal Ordering.
$\xi$ is in units of $\pi$.
The contours are not continued to the right of $k^2=0.9$ where the validity of Eq.~\eqref{nosc} becomes problematic as $k^2\to 1$.
We expect a more thorough analysis to lead to the contours dropping down to the $k^2$ axis near $k^2=1$.
}
\label{AtmosBounds}
\end{figure}

The expression we obtained in Eq.~\eqref{nosc} is just the usual neutrino oscillation 
formula scaled by the constant $\left(c_\xi^2+s_\xi^2\frac{k^2}{2}\right)$. 
If we consider atmospheric neutrino oscillations,
this factor cannot always be absorbed into $\sin^2{(2\theta_{23})}\leq1$, so we can interpret the bounds on the quantity $\sin^2{(2\theta_{23})}$ as bounds on $\left(c_\xi^2+s_\xi^2\frac{k^2}{2}\right)\;\sin^2{(2\theta_{23})}$.  
According to TABLE~III of Ref.~\cite{deSalas:2020pgw}, the current bounds on $\sin^2\theta_{23}$ are
\begin{eqnarray}
\sin^2\theta_{23}
& = & 0.566{}^{+0.016}_{-0.022} \qquad\;\;\;\mbox{(Normal Ordering, $1\sigma$)} \vphantom{\Big|}\cr
& = & 0.505 - 0.596 \qquad \mbox{(NO, $2\sigma$)}\;, \vphantom{\Big|}\cr
& = & 0.441 - 0.609 \qquad \mbox{(NO, $3\sigma$)}\;, \vphantom{\Big|}\cr
& = & 0.566{}^{+0.018}_{-0.023} \qquad\;\;\;\mbox{(Inverted Ordering, $1\sigma$)} \vphantom{\Big|}\cr
& = & 0.514 - 0.597 \qquad \mbox{(IO, $2\sigma$)}\;, \vphantom{\Big|}\cr
& = & 0.446 - 0.609 \qquad \mbox{(IO, $3\sigma$)}\;, \vphantom{\Big|}
\end{eqnarray}
leading to the bounds on $\sin^2 2\theta_{23} = 4\sin^2\theta_{23}(1-\sin^2\theta_{23})$ given by
\begin{eqnarray}
\sin^2 2\theta_{23}
& = & 0.973 - 0.992 \qquad\mbox{(NO, $1\sigma$)} \vphantom{\big|}\cr
& = & 0.963 - 1.000 \qquad\mbox{(NO, $2\sigma$)} \vphantom{\big|}\cr
& = & 0.952 - 1.000 \qquad\mbox{(NO, $3\sigma$)} \vphantom{\big|}\cr
& = & 0.972 - 0.993 \qquad\mbox{(IO, $1\sigma$)} \vphantom{\big|}\cr
& = & 0.962 - 0.999 \qquad\mbox{(IO, $2\sigma$)} \vphantom{\big|}\cr
& = & 0.952 - 1.000 \qquad\mbox{(IO, $3\sigma$)} \vphantom{\big|}
\end{eqnarray}
So the bounds for both the NO and IO cases are more or less the same.
%
%
%
The NO numbers translate to 
\begin{eqnarray}
s_\xi^2\left(1-\dfrac{k^2}{2}\right) 
& < & 1 - 0.973 \;=\; 0.027 \qquad(1\sigma)\;,\cr
& < & 1 - 0.963 \;=\; 0.037 \qquad(2\sigma)\;,\cr
& < & 1 - 0.952 \;=\; 0.048 \qquad(3\sigma)\;.\vphantom{\bigg|}
\end{eqnarray}
For the $k=0$ case, this translates to
\begin{equation}
\dfrac{\xi}{\pi} \;<\;
0.053\;\;(1\sigma)\;,\quad
0.062\;\;(2\sigma)\;,\quad
0.070\;\;(3\sigma)\;.
\label{mzeroAtmosNuBounds}
\end{equation}
For non-zero $k$, see FIG.~\ref{AtmosBounds} for the likelihood contours on the $k^2$-$\xi$ plane. 
In FIG.~\ref{AtmosBounds}, the contours are not extended all the way to $k^2=1$ since Eq.~\eqref{nosc} is unreliable 
in that region. 
They can be expected to turn downward and reach the $k^2$ axis before reaching $k^2=1$ since
the oscillation function $\overline{\mathcal{F}}(\Delta)$ will start to deviate from the sinusoidal as $k^2\to 1$ as shown in FIG.~\ref{NQMoscillations}.
A complete oscillation analysis would be necessary to work out the bounds on $\xi$ and $k^2$ in this region.
We do not perform this analysis in this paper due to reasons that are given in the
discussion session.
Instead, we look at $B^0$-$\overline{B^0}$ oscillation next.

\section{$B^0$-$\overline{B^0}$ Oscillation}\label{BBarBounds}

Another oscillation phenomenon that has been extensively studied in experiments is that of time-dependent CP asymmetry in meson-antimeson oscillations \cite{Asner:2022,Schneider:2022} for which 
a wealth of data exist from Belle and BABAR. 
In the following, we place bounds on $k$ and $\xi$
utilizing the $B^0$-$\overline{B^0}$ oscillation data from Belle.

\subsection{Canonical QM Case}

Consider the oscillation of $B^0$ into $\overline{B^0}$ and vice versa before they decay into a common physical state $f$. 
Recall that in the usual canonical QM formulation,
the CP eigenstates are
$(\ket{B^0}\pm\ket{\overline{B^0}})/\sqrt{2}$,
whereas the mass eigenstates are \cite{Bigi:2000yz}
\begin{eqnarray}
\ket{B_L^0} & = & p\ket{B^0} + q\ket{\overline{B^0}} \;,
\vphantom{\Big|}\cr
\ket{B_H^0} & = & p\ket{B^0} - q\ket{\overline{B^0}} \;,
\vphantom{\Big|}
\end{eqnarray}
where $|p| = |q| = 1/\sqrt{2}$, and $p$ and $q$ have opposite phase:
\begin{equation}
p \;=\; \dfrac{1}{\sqrt{2}}\,\vec{x}(\phi)\;,\qquad
q \;=\; \dfrac{1}{\sqrt{2}}\,\vec{x}(-\phi)\;.
\end{equation}
To leading order in the Standard Model, 
the CP violating phase $\phi$ is given by \cite{Bigi:2000yz}
\begin{equation}
\vec{x}(2\phi)\;=\;
\dfrac{p}{q}\;=\;
\frac{V_{tb}^{\vphantom{*}}V^*_{td}}
{V^*_{tb}V_{td}^{\vphantom{*}}} 
\;.
\end{equation}
To facilitate the application of our Nambu extension later, we work in the mass eigenstate basis, 
in which the $B^0$ and $\overline{B^0}$ states are expressed as superpositions of the mass eigenstates $B_L^0$ and $B_H^0$ :
\begin{eqnarray}
\ket{B^0} & = & 
q\ket{B^0_L} + 
q\ket{B^0_H} 
\;,
\vphantom{\Big|}\cr
\ket{\overline{B^0}} & = & 
p\ket{B^0_L} - 
p\ket{B^0_H} 
\;.
\vphantom{\Big|}
\end{eqnarray}
In the rest frame of the particle,
the mass eigenstates
$\ket{B^0_L}$ and $\ket{B^0_H}$ evolve as
\begin{equation}
\ket{B_L^0(t)}\,=\,e^{-(\Gamma_L/2)t}\,\vec{x}(-m_L t)\ket{B_L^0(0)}\;,\qquad
\ket{B_H^0(t)}\,=\,e^{-(\Gamma_H/2)t}\,\vec{x}(-m_H t)\ket{B_H^0(0)}\;,
\end{equation}
so the $B^0$ superposition evolves as
\begin{eqnarray}
\ket{B^0(0)}
& = &
\dfrac{1}{\sqrt{2}}\vec{x}(-\phi)\ket{B^0_L(0)} + 
\dfrac{1}{\sqrt{2}}\vec{x}(-\phi)\ket{B^0_H(0)}
\cr
& \downarrow & \cr
\ket{B^0(t)}
& = &
\dfrac{1}{\sqrt{2}}\,
e^{-(\Gamma_L/2)t}\,\vec{x}(-\phi)\ket{B^0_L(t)} 
+ \dfrac{1}{\sqrt{2}}\,e^{-(\Gamma_H/2)t}\,\vec{x}(-\phi)\ket{B^0_H(t)}
\vphantom{\Bigg|}\cr
& = &
\dfrac{1}{\sqrt{2}}\,
e^{-(\Gamma_L/2)t}\,\vec{x}(-\phi-m_L t)\ket{B^0_L(0)} + \dfrac{1}{\sqrt{2}}\,
e^{-(\Gamma_H/2)t}\,
\vec{x}(-\phi-m_H t)\ket{B^0_H(0)}
\;.
\end{eqnarray}
We assume that $\Gamma_L=\Gamma_H$, and
drop the common decay factor from this point on.
Then, we find
\begin{eqnarray}
g(B^0(0),B^0(t))
& = &
 \dfrac{1}{2}\vec{x}(-\phi)\cdot\vec{x}(-\phi-m_L t)
+\dfrac{1}{2}\vec{x}(-\phi)\cdot\vec{x}(-\phi-m_H t)
\vphantom{\Bigg|}\cr
& = & \dfrac{1}{2}
\Big[
\cos(m_L t) + \cos(m_H t)
\Big]
\vphantom{\Bigg|}\cr
& = & \cos(mt) \cos\bigg(\dfrac{\Delta m\,t}{2}\bigg)
\;,
\vphantom{\bigg|}
\cr
\varepsilon(B^0(0),B^0(t))
& = &
 \dfrac{1}{2}\vec{x}(-\phi)\times\vec{x}(-\phi-m_L t)
+\dfrac{1}{2}\vec{x}(-\phi)\times\vec{x}(-\phi-m_H t)
\vphantom{\Bigg|}\cr
& = & - \dfrac{1}{2}\Big[
\sin(m_L t) + \sin(m_H t)
\Big]
\vphantom{\Bigg|}\cr
& = & -\sin(mt) \cos\bigg(\dfrac{\Delta m\,t}{2}\bigg)
\;,
\vphantom{\bigg|}
\cr
g(\overline{B^0}(0),B^0(t))
& = &
 \dfrac{1}{2}\vec{x}(\phi)\cdot\vec{x}(-\phi-m_L t)
-\dfrac{1}{2}\vec{x}(\phi)\cdot\vec{x}(-\phi-m_H t)
\vphantom{\Bigg|}\cr
& = & \dfrac{1}{2}
\Big[
\cos(2\phi + m_L t) - \cos(2\phi + m_H t)
\Big]
\vphantom{\Bigg|}\cr
& = & \sin(2\phi + mt) \sin\bigg(\dfrac{\Delta m\,t}{2}\bigg)
\;,
\vphantom{\bigg|}
\cr
\varepsilon(\overline{B^0}(0),B^0(t))
& = &
 \dfrac{1}{2}\vec{x}(\phi)\times\vec{x}(-\phi-m_L t)
-\dfrac{1}{2}\vec{x}(\phi)\times\vec{x}(-\phi-m_H t)
\vphantom{\Bigg|}\cr
& = & -\dfrac{1}{2}
\Big[
\sin(2\phi + m_L t) - \sin(2\phi + m_H t)
\Big]
\vphantom{\Bigg|}\cr
& = & \cos(2\phi + mt) \sin\bigg(\dfrac{\Delta m\,t}{2}\bigg)
\;,
\vphantom{\bigg|}
\end{eqnarray}
where
\begin{equation}
m \,=\, \dfrac{m_L + m_H}{2}\;,\qquad
\Delta m \,=\, m_H - m_L\;.
\end{equation}
The survival and transition probabilities evolve with
time as
\begin{eqnarray}
P(B^0\to B^0)
& = &  g(B^0(0),B^0(t))^2 + \varepsilon(B^0(0),B^0(t))^2 
\,=\, \cos^2\left(\dfrac{\Delta m\,t}{2}\right)
\;,\vphantom{\Bigg|}\cr
P(B^0\to \overline{B^0})
& = & g(\overline{B^0}(0),B^0(t))^2 + \varepsilon(\overline{B^0}(0),B^0(t))^2 
\,=\, \sin^2\left(\dfrac{\Delta m\,t}{2}\right)
\;,\vphantom{\Bigg|}
\end{eqnarray}
with similar relations for
$P(\overline{B^0}\to B^0)$ and
$P(\overline{B^0}\to \overline{B^0})$.

Consider the asymmetry 
\begin{equation}
A(t)\;=\;
\dfrac{|\braket{f}{\overline{B^0}(t)}|^2-|\braket{f}{B^0(t)}|^2}
      {|\braket{f}{\overline{B^0}(t)}|^2+|\braket{f}{B^0(t)}|^2}
\;,
\label{Atdef}
\end{equation}
where $\braket{f}{B^0(t)}$ and $\braket{f}{\overline{B}^0(t)}$ are the amplitudes of 
the states 
$\ket{B^0(t)}$ and $\ket{\overline{B}^0(t)}$ decaying into $f$ respectively.
Let
\begin{equation}
\braket{f}{B^0_L(0)}\,=\, A_L\,\vec{x}(-\lambda)\;,\qquad
\braket{f}{B^0_H(0)}\,=\, A_H\,\vec{x}(-\eta)\;,
\end{equation}
that is
\begin{equation}
\ket{f}
\;=\;
\ket{B^0_L(0)}\underbrace{\braket{B^0_L(0)}{f}}_{\displaystyle A_L\vec{x}(\lambda)} +
\ket{B^0_H(0)}\underbrace{\braket{B^0_H(0)}{f}}_{\displaystyle A_H\vec{x}(\eta)}
+\cdots 
\;.
\end{equation}
Then,
\begin{eqnarray}
g(f,B^0(t))
& = & \dfrac{A_L}{\sqrt{2}}
\,\vec{x}(\lambda)\cdot\vec{x}(-\phi-m_L t)
+ \dfrac{A_H}{\sqrt{2}}
\,\vec{x}(\eta)\cdot\vec{x}(-\phi-m_H t)
\vphantom{\Bigg|}\cr
& = & \dfrac{A_L}{\sqrt{2}}
\cos(\lambda+\phi+m_L t)
+ \dfrac{A_H}{\sqrt{2}}
\cos(\eta+\phi+m_H t)
\;,\vphantom{\Bigg|}\cr
\varepsilon(f,B^0(t))
& = &
\dfrac{A_L}{\sqrt{2}}
\,\vec{x}(\lambda)\times\vec{x}(-\phi-m_L t)
+ \dfrac{A_H}{\sqrt{2}}
\,\vec{x}(\eta)\times\vec{x}(-\phi-m_H t)
\vphantom{\Bigg|}\cr
& = &
-\dfrac{A_L}{\sqrt{2}}
\sin(\lambda+\phi+m_L t)
-\dfrac{A_H}{\sqrt{2}}
\sin(\eta+\phi+m_H t)
\;,\vphantom{\Bigg|}
\vphantom{\Bigg|}
\end{eqnarray}
and
\begin{eqnarray}
|\braket{f}{B^0(t)}|^2
& = & g(f,B^0(t))^2 + \varepsilon(f,B^0(t))^2 
\vphantom{\bigg|}\cr
& = & 
\dfrac{A_L^2}{2} + \dfrac{A_H^2}{2}
+ A_L A_H 
\Big\{
 \cos(\eta-\lambda)\cos(\Delta m\,t)
-\sin(\eta-\lambda)\sin(\Delta m\,t)
\Big\}
\;.
\end{eqnarray}
Similarly,
\begin{equation}
|\braket{f}{\overline{B^0}(t)}|^2
\;=\;
\dfrac{A_L^2}{2} + \dfrac{A_H^2}{2}
- A_L A_H 
\Big\{
 \cos(\eta-\lambda)\cos(\Delta m\,t)
-\sin(\eta-\lambda)\sin(\Delta m\,t)
\Big\}
\;,
\end{equation}
and we find
\begin{eqnarray}
A(t)
& = &
 \underbrace{\dfrac{2A_L A_H}{A_L^2+A_H^2}\sin(\eta-\lambda)}_{\displaystyle S}\sin(\Delta m\,t)
-\underbrace{\dfrac{2A_L A_H}{A_L^2+A_H^2}\cos(\eta-\lambda)}_{\displaystyle C}\cos(\Delta m\,t)
\cr
& = & S\sin(\Delta m\,t) - C\cos(\Delta m\,t)
\vphantom{\bigg|}\;.
\label{AtCanonialQM}
\end{eqnarray}
Note that
\begin{equation}
S^2 + C^2 
\;=\; 
\bigg(\dfrac{2A_L A_H}{A_L^2+A_H^2}\bigg)^2
\;\le\; 1\;.
\label{S2C2}
\end{equation}
%
%
It is straightforward to show that the above definitions of $S$ and $C$ agree with the usual 
expressions \cite{Bigi:2000yz} involving
\begin{eqnarray}
A_f
& = & q\Big(\braket{f}{B_L}+\braket{f}{B_H}\Big)
\;=\;\braket{f}{B^0}\;,\quad\text{and}
\vphantom{\bigg|}\cr
\bar{A}_f
& = & p\Big(\braket{f}{B_L}-\braket{f}{B_H}\Big)
\;=\;\braket{f}{\overline{B}^0}
\;.\vphantom{\bigg|}
\end{eqnarray}
We refrain from this rewriting, however, since the analogs of $A_f$ and $\bar{A}_f$ are ill defined in Nambu QM.

\subsection{Nambu QM Extension}

Let us now apply the Nambu extension to the above formalism.
We assume that the $B^0$ and $\overline{B^0}$ states are given by the superpositions
\begin{eqnarray}
\ket{B^0} 
& = & Q\ket{B^0_L} + Q\ket{B^0_H} \;,
\vphantom{\Big|}\cr
\ket{\overline{B^0}} 
& = & P\ket{B^0_L} - P\ket{B^0_H} \;,
\vphantom{\Big|}
\label{B0B0barStates}
\end{eqnarray}
where
\begin{equation}
P\,=\,\dfrac{1}{2Q}\,=\,\dfrac{1}{\sqrt{2}}\,\vec{X}( \phi)\;,\qquad
Q\,=\,\dfrac{1}{2P}\,=\,\dfrac{1}{\sqrt{2}}\,\vec{X}(-\phi)\;.
\end{equation}
Here, we suppress the dependence of $\vec{X}$ on the deformation parameters $k$ and $\xi$ to
simplify our notation.

In the rest frame of the particle, the mass eigenstates
$\ket{B^0_L}$ and $\ket{B^0_H}$ evolve as
\begin{equation}
\ket{B_L^0(t)}\,=\,e^{-(\Gamma_L/2)t}\,\vec{X}(-m_L t)\ket{B_L^0(0)}\;,\qquad
\ket{B_H^0(t)}\,=\,e^{-(\Gamma_H/2)t}\,\vec{X}(-m_H t)\ket{B_H^0(0)}\;.
\end{equation}
We assume $\Gamma_L=\Gamma_H$ as in the canonical case and suppress this common decay factor from this point on.
The state $\ket{B^0(0)} = \ket{B^0}$ evolves to
\begin{eqnarray}
\ket{B^0(t)}
& = &
\dfrac{1}{\sqrt{2}}\,
\vec{X}(-\phi-m_L t)\ket{B^0_L(0)} + 
\dfrac{1}{\sqrt{2}}\,
\vec{X}(-\phi-m_H t)\ket{B^0_H(0)}
\;.
\end{eqnarray}
Therefore,
\begin{eqnarray}
g(B^0(0),B^0(t))
& = &
 \dfrac{1}{2}\,\vec{X}(-\phi)\cdot\vec{X}(-\phi-m_L t)
+\dfrac{1}{2}\,\vec{X}(-\phi)\cdot\vec{X}(-\phi-m_H t)
\vphantom{\Bigg|}\cr
& = & \dfrac{1}{2}\bigg[
c_\xi^2\,\Cn_\phi\,\Big(\Cn_L + \Cn_H\Big)
+(c_\xi^2+s_\xi^2 k^2)\,\Sn_\phi\,\Big(\Sn_L + \Sn_H\Big)
+s_\xi^2\,\Dn_\phi\,\Big(\Dn_L + \Dn_H\Big)
\bigg]
\;,
\vphantom{\bigg|}\cr
\vec{\varepsilon}\,(B^0(0),B^0(t))
& = &
 \dfrac{1}{2}\,\vec{X}(-\phi)\times\vec{X}(-\phi-m_L t)
+\dfrac{1}{2}\,\vec{X}(-\phi)\times\vec{X}(-\phi-m_H t)
\vphantom{\Bigg|}\cr
& = & \dfrac{1}{2}
\begin{bmatrix}
s_\xi\sqrt{c_\xi^2+s_\xi^2 k^2}
\Big\{\Sn_\phi\Big(\Dn_L + \Dn_H\Big)-\Dn_\phi\Big(\Sn_L + \Sn_H\Big)\Big\}
\vphantom{\bigg|}\\
-s_\xi c_\xi 
\Big\{\Dn_\phi\Big(\Cn_L + \Cn_H\Big)-\Cn_\phi\Big(\Dn_L + \Dn_H\Big)\Big\}
\vphantom{\bigg|}\\
-c_\xi\sqrt{c_\xi^2+s_\xi^2 k^2}
\Big\{\Cn_\phi\Big(\Sn_L + \Sn_H\Big)-S_\phi\Big(\Cn_L + \Cn_H\Big)\Big\}
\vphantom{\bigg|}
\end{bmatrix}
,\cr
& & 
\vphantom{\Big|}\cr
g(\overline{B^0}(0),B^0(t))
& = &
 \dfrac{1}{2}\,\vec{X}(\phi)\cdot\vec{X}(-\phi-m_L t)
-\dfrac{1}{2}\,\vec{X}(\phi)\cdot\vec{X}(-\phi-m_H t)
\vphantom{\Bigg|}\cr
& = & \dfrac{1}{2}\bigg[
c_\xi^2\,\Cn_\phi\,\Big(\Cn_L - \Cn_H\Big)
-(c_\xi^2+s_\xi^2 k^2)\,\Sn_\phi\,\Big(\Sn_L - \Sn_H\Big)
+s_\xi^2\,\Dn_\phi\,\Big(\Dn_L - \Dn_H\Big)
\bigg]
\;,
\vphantom{\bigg|}\cr
\vec{\varepsilon}\,(\overline{B^0}(0),B^0(t))
& = &
 \dfrac{1}{2}\,\vec{X}(\phi)\times\vec{X}(-\phi-m_L t)
-\dfrac{1}{2}\,\vec{X}(\phi)\times\vec{X}(-\phi-m_H t)
\vphantom{\Bigg|}\cr
& = & \dfrac{1}{2}
\begin{bmatrix}
-s_\xi\sqrt{c_\xi^2+s_\xi^2 k^2}
\Big\{\Sn_\phi\Big(\Dn_L - \Dn_H\Big)-\Dn_\phi\Big(\Sn_L - \Sn_H\Big)\Big\}
\vphantom{\bigg|}\\
-s_\xi c_\xi 
\Big\{\Dn_\phi\Big(\Cn_L - \Cn_H\Big)-\Cn_\phi\Big(\Dn_L - \Dn_H\Big)\Big\}
\vphantom{\bigg|}\\
-c_\xi\sqrt{c_\xi^2+s_\xi^2 k^2}
\Big\{\Cn_\phi\Big(\Sn_L - \Sn_H\Big)-S_\phi\Big(\Cn_L - \Cn_H\Big)\Big\}
\vphantom{\bigg|}
\end{bmatrix}
\;,
\end{eqnarray}
where we have used the shorthand
\begin{equation}
\begin{array}{rlrlrl}
\Cn_\phi &\equiv\, \Cn(\phi)\;, &
\qquad \Cn_L &\equiv\, \Cn(\phi+m_L t)\;, &
\qquad \Cn_H &\equiv\, \Cn(\phi+m_H t)\;,\\
\Sn_\phi &\equiv\, \Sn(\phi)\;, &
\qquad \Sn_L &\equiv\, \Sn(\phi+m_L t)\;, &
\qquad \Sn_H &\equiv\, \Sn(\phi+m_H t)\;,\\
\Dn_\phi &\equiv\, \Dn(\phi)\;, &
\qquad \Dn_L &\equiv\, \Dn(\phi+m_L t)\;, &
\qquad \Dn_H &\equiv\, \Dn(\phi+m_H t)\;.
\end{array}
\end{equation}
The transition probability is given by
\begin{eqnarray}
P(B^0\to \overline{B^0})
& = & g(\overline{B^0}(0),B^0(t))^2 + \vec{\varepsilon}\,(\overline{B^0}(0),B^0(t))\cdot\vec{\varepsilon}\,(\overline{B^0}(0),B^0(t))
\vphantom{\bigg|}\cr
& = & \dfrac{1 - \big\{ c_\xi^2(\Cn_L\Cn_H + \Sn_L\Sn_H) + s_\xi^2(k^2\Sn_L\Sn_H + \Dn_L\Dn_H)\big\}}{2}
\vphantom{\Bigg|}\cr
& = &
\dfrac{(c_\xi^2 + s_\xi^2 k^2) - k^2\Sn^2(\phi+mt)}
{1-k^2\,\Sn^2 (\phi+mt)\,\Sn^2(\Delta m\,t/2)}
\;\Sn^2(\Delta m\,t/2)
\vphantom{\Bigg|}\cr
& = &
\mathcal{F}(\phi+mt,\Delta m t)
\;,\vphantom{\Bigg|}
\end{eqnarray} 
where the function $\mathcal{F}$
was defined in Eq.~\eqref{Fdef},
and $m = (m_L + m_H)/2$ and $\Delta m = m_H-m_L$.
The survival probably is also
found to be
\begin{eqnarray}
P(B^0\to B^0)
& = & g(B^0(0),B^0(t))^2 + \vec{\varepsilon}\,(B^0(0),B^0(t))\cdot\vec{\varepsilon}\,(B^0(0),B^0(t))
\vphantom{\bigg|}\cr
& = & 1 - \mathcal{F}(\phi+mt,\Delta m t)
\;.\vphantom{\Bigg|}
\end{eqnarray} 
Note these expressions are the same as the neutrino oscillation case we considered earlier, Eq.~\eqref{PAAandPAB}, with mixing angle set to $\pi/4$.
The CP violating phase $\phi$ only appears as a shift of the
first argument of $\mathcal{F}$.

The mass difference between 
$B^0_L$ and $B^0_H$ is
\cite{ParticleDataGroup:2022pth,HeavyFlavorAveragingGroup:2022wzx}
\begin{eqnarray}
\Delta m
\;=\; m_H - m_L
& = & 0.5065\pm 0.0019\;\hbar\cdot\mathrm{ps}^{-1}
\vphantom{\Big|}\cr
& = & (3.334\pm 0.013)\times 10^{-4}\,\mathrm{eV}
\;,\vphantom{\Big|}
\label{PDBDeltam}
\end{eqnarray}
whereas the average mass is \cite{ParticleDataGroup:2022pth,HeavyFlavorAveragingGroup:2022wzx}
\begin{equation}
m \;=\;
5279\pm 0.12\;\mathrm{MeV}\;.
\end{equation}
Therefore,
\begin{equation}
\dfrac{2m}{\Delta m}
\;=\;
\dfrac{m_H+m_L}
      {m_H-m_L}
\;\approx\;
3\times 10^{13}\;,
\end{equation}
and we conclude that $\Sn^2(\phi + mt)$ oscillates very rapidly compared to $\Sn^2(\Delta m\,t)$ and will average out in $\mathcal{F}(\phi+mt,\Delta m\,t)$,
just as in the neutrino oscillation case.  So the transition probability
is well approximated by
\begin{equation}
P(B^0\to \overline{B^0})
\;\approx\; \overline{\mathcal{F}}(\Delta m\, t)
\;\approx\;
\bigg(c_\xi^2 + s_\xi^2\dfrac{k^2}{2}\bigg)\sin^2\left(\dfrac{\Delta m\,t}{2}\right)\;,
\end{equation}
where the rightmost expression is valid when $k^2$ is not too close to one.

\begin{figure}[ht]
\subfigure[$\cos\Theta_1$]{\includegraphics[height=5cm]{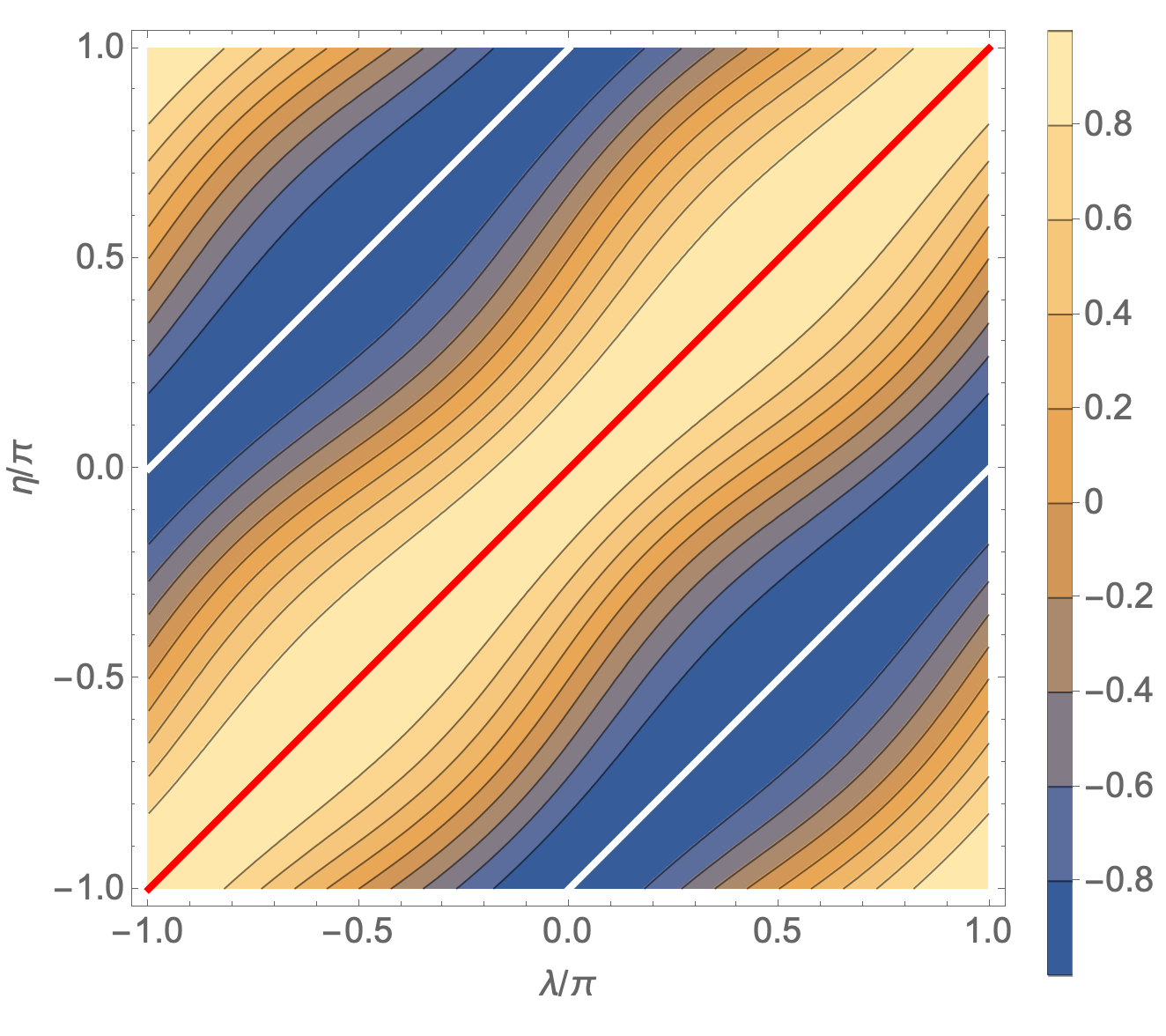}}
\subfigure[$\sin\Theta_1$]{\includegraphics[height=5cm]{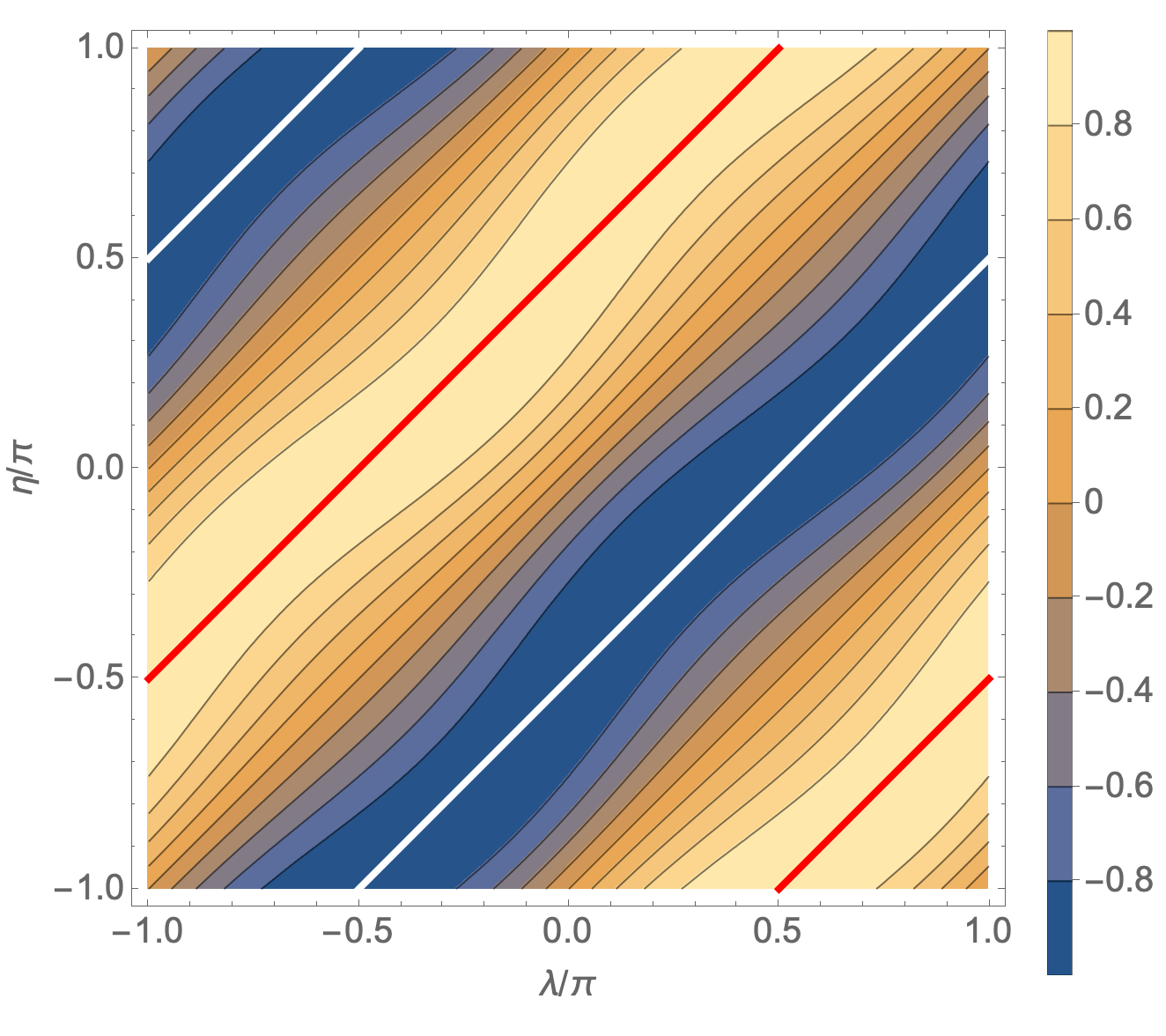}}
\subfigure[$\cos\Theta_2$]{\includegraphics[height=5cm]{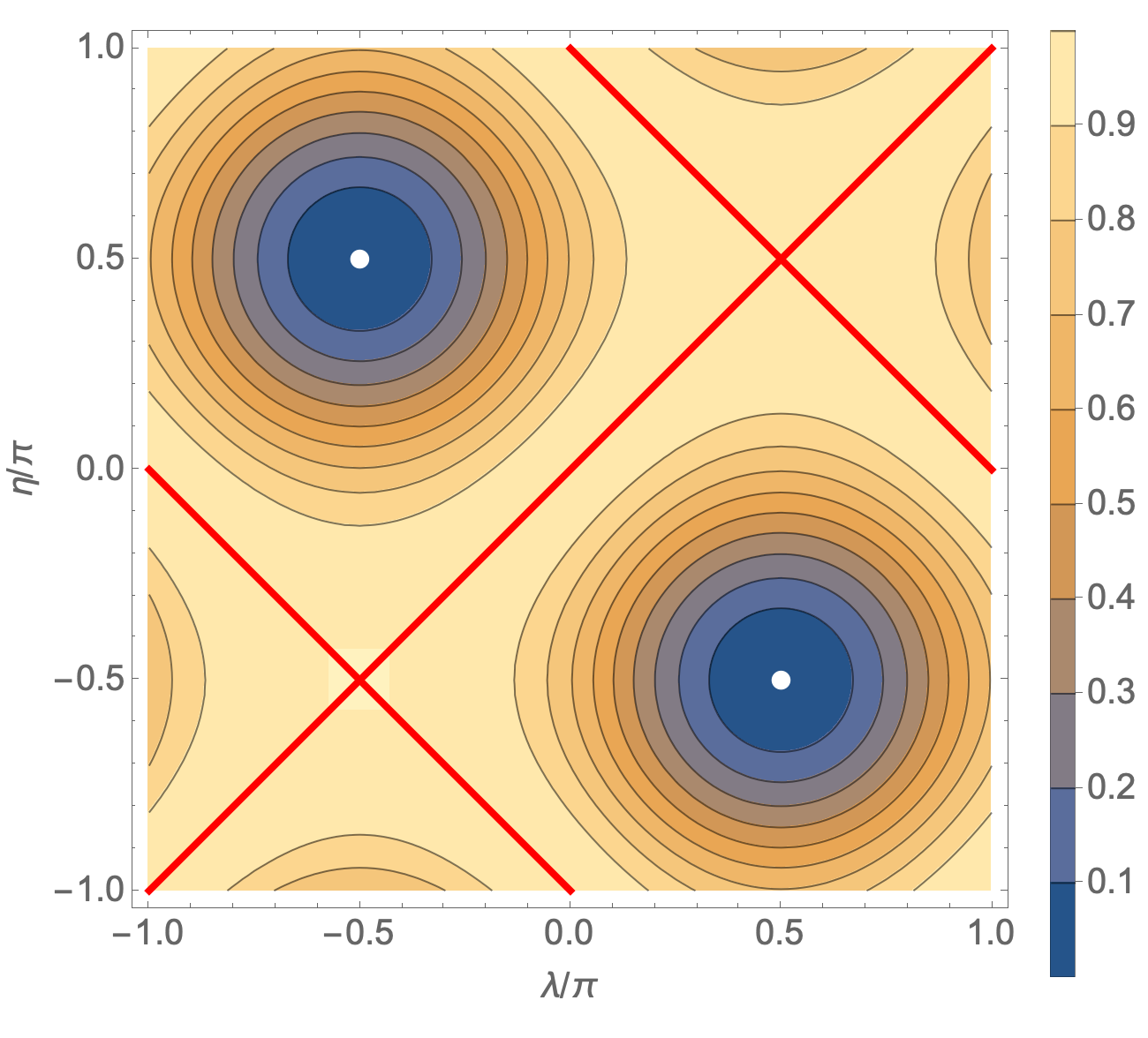}}
\caption{Contour plots of 
(a) $\cos\Theta_1$,
(b) $\sin\Theta_1$, and
(c) $\cos\Theta_2$ for the case $k^2=0.5$ on the $(\lambda,\eta)$ plane
for a single period in each direction.  The thick red 
lines on the graphs indicate the points at which each function is at their maxima,
while the white lines on the $\cos\Theta_1$ graph and the while dots on the $\cos\Theta_2$ graph indicate points at which
they are at their respective minima.
}
\label{Theta1Theta2Plot}
\end{figure}

Let us now consider the asymmetry $A(t)$ defined in Eq.~\eqref{Atdef}.
We assume
\begin{equation}
\braket{f}{B^0_L(0)}\,=\, A_L\vec{X}(-\lambda)\;,\qquad
\braket{f}{B^0_H(0)}\,=\, A_H\vec{X}(-\eta)\;.
\end{equation}
Then
\begin{eqnarray}
g(f,B^0(t))
& = & \dfrac{A_L}{\sqrt{2}}
\,\vec{X}(\lambda)\cdot\vec{X}(-\phi-m_L t)
+ \dfrac{A_H}{\sqrt{2}}
\,\vec{X}(\eta)\cdot\vec{X}(-\phi-m_H t)
\vphantom{\Bigg|}\cr
& = & \dfrac{A_L}{\sqrt{2}}
\Big[
c_\xi^2\Cn_\lambda\Cn_L - (c_\xi^2+s_\xi^2 k^2)\Sn_\lambda\Sn_L + s_\xi^2\Dn_\lambda\Dn_L
\Big]
\cr
& & 
+ \dfrac{A_H}{\sqrt{2}}
\Big[
c_\xi^2\Cn_\eta\Cn_H - (c_\xi^2+s_\xi^2 k^2)\Sn_\eta\Sn_H + s_\xi^2\Dn_\eta\Dn_H
\Big]
\;,\vphantom{\Bigg|}\cr
\vec{\varepsilon}\,(f,B^0(t))
& = &
\dfrac{A_L}{\sqrt{2}}
\,\vec{X}(\lambda)\times\vec{X}(-\phi-m_L t)
+ \dfrac{A_H}{\sqrt{2}}
\,\vec{X}(\eta)\times\vec{X}(-\phi-m_H t)
\vphantom{\Bigg|}\cr
& = &
-\dfrac{A_L}{\sqrt{2}}
\begin{bmatrix}
s_\xi\sqrt{c_\xi^2+s_\xi^2 k^2}\Big(\Sn_\lambda\Dn_L + \Dn_\lambda\Sn_L\Big) \vphantom{\bigg|} \\
s_\xi c_\xi\Big(\Dn_\lambda\Cn_L - \Cn_\lambda\Dn_L\Big) \vphantom{\bigg|} \\
c_\xi\sqrt{c_\xi^2+s_\xi^2 k^2}\Big(\Cn_\lambda\Sn_L + \Sn_\lambda\Cn_L\Big) \vphantom{\bigg|}
\end{bmatrix}
- \dfrac{A_H}{\sqrt{2}}
\begin{bmatrix}
s_\xi\sqrt{c_\xi^2+s_\xi^2 k^2}\Big(\Sn_\eta\Dn_H + \Dn_\eta\Sn_H\Big) \vphantom{\bigg|} \\
s_\xi c_\xi\Big(\Dn_\eta\Cn_H - \Cn_\eta\Dn_H\Big) \vphantom{\bigg|} \\
c_\xi\sqrt{c_\xi^2+s_\xi^2 k^2}\Big(\Cn_\eta\Sn_H + \Sn_\eta\Cn_H\Big) \vphantom{\bigg|}
\end{bmatrix}
,
\cr
& &
\end{eqnarray}
and we find
\begin{eqnarray}
\lefteqn{
|\braket{f}{B^0(t)}|^2}
\cr
& = & g(f,B^0(t))^2 + \vec{\varepsilon}\,(f,B^0(t))\cdot\vec{\varepsilon}\,(f,B^0(t))
\vphantom{\bigg|}\cr
& = & \dfrac{A_L^2}{2} + \dfrac{A_H^2}{2}
\cr
& & + A_L A_H
\bigg[
\Big\{c_\xi^2
\underbrace{(\Cn_\lambda\Cn_\eta + \Sn_\lambda\Sn_\eta)}_{\displaystyle \cos\Theta_1}
+s_\xi^2 
\underbrace{(k^2\Sn_\lambda\Sn_\eta + \Dn_\lambda\Dn_\eta)}_{\displaystyle \cos\Theta_2}
\Big\}
\cr
& & \qquad\qquad\times
\Big\{c_\xi^2
(\Cn_L\Cn_H + \Sn_L\Sn_H)
+s_\xi^2 
(k^2\Sn_L\Sn_H + \Dn_L\Dn_H)
\Big\}
\vphantom{\Bigg|}\cr
& & \qquad\qquad
+c_\xi^2(c_\xi^2 + s_\xi^2 k^2)
\underbrace{\Big(
\Sn_\lambda\Cn_\eta - \Cn_\lambda\Sn_\eta
\Big)}_{\displaystyle -\sin\Theta_1}
\Big(
\Sn_H\Cn_L - \Cn_H\Sn_L
\Big)
\vphantom{\bigg|}\cr
& & \qquad\qquad
+s_\xi^2(c_\xi^2 + s_\xi^2 k^2)
\underbrace{
\Big(
\Sn_\lambda\Dn_\eta - \Dn_\lambda\Sn_\eta
\Big)
}_{\displaystyle \big[\Sn,\Dn\big]_{\lambda,\eta}
}
\Big(
\Sn_H\Dn_L - \Dn_H\Sn_L
\Big)
\vphantom{\bigg|}\cr
& & \qquad\qquad
-s_\xi^2 c_\xi^2
\underbrace{
\Big(
\Cn_\lambda\Dn_\eta - \Dn_\lambda\Cn_\eta
\Big)
}_{\displaystyle
\big[\Cn,\Dn\big]_{\lambda,\eta}
}
\Big(
\Cn_H\Dn_L - \Dn_H\Cn_L
\Big)
\bigg]
\vphantom{\Bigg|}\cr
& = & \dfrac{A_L^2}{2} + \dfrac{A_H^2}{2}
\cr
& & + A_L A_H \Bigg[
\Big\{
c_\xi^2\cos\Theta_1 + s_\xi^2\cos\Theta_2
\Big\}
\vphantom{\dfrac{\Big|}{\Big|}}\cr
& & \qquad\qquad\quad\times
\underbrace{
\left\{c_\xi^2\left(1-\dfrac{2\;\Dn^2\left(\phi+mt\right)\Sn^2\left(\frac{\Delta m\,t}{2}\right)}{1-k^2\,\Sn^2\left(\phi+mt\right)\Sn^2\left(\frac{\Delta m\,t}{2}\right)}\right)+s_\xi^2\left(1-\frac{2k^2\;\Cn^2\left(\phi+mt\right)\Sn^2\left(\frac{\Delta m\,t}{2}\right)}{1-k^2\,\Sn^2\left(\phi+mt\right)\Sn^2\left(\frac{\Delta m\,t}{2}\right)}\right)
\right\}
}_{\displaystyle
1-2\mathcal{F}(\phi+mt,\Delta m \,t)}
\vphantom{\dfrac{\Big|}{\Big|}}\cr
& & \qquad\qquad\quad
-\;c_\xi^2
\left(c_\xi^2+k^2\,s_\xi^2\right)
\sin\Theta_1
\underbrace{
\left(
\dfrac{2\,\Dn(\phi+mt)\,\Cn\left(\tfrac{\Delta m\,t}{2}\right)\Sn\left(\tfrac{\Delta m\,t}{2}\right)}
{1-k^2\,\Sn^2\left(\phi+mt\right)\Sn^2\left(\frac{\Delta m\,t}{2}\right)}
\right)
}_{\displaystyle
\equiv \mathcal{G}(\phi+mt,\Delta m\,t)
}
\vphantom{\dfrac{\Big|}{\Big|}}\cr
& & \qquad\qquad\quad
+\;s_\xi^2\left(c_\xi^2+k^2\,s_\xi^2\right)
\big[\Sn,\Dn\big]_{\lambda,\eta}
\underbrace{
\left(
\dfrac{2\,\Cn(\phi+mt)\,\Dn\left(\tfrac{\Delta m\,t}{2}\right)\Sn\left(\tfrac{\Delta m\,t}{2}\right)}
{1-k^2\,\Sn^2\left(\phi+mt\right)\Sn^2\left(\frac{\Delta m\,t}{2}\right)}
\right)
}_{\displaystyle \equiv\mathcal{H}(\phi+mt, \Delta m\,t)}
\vphantom{\dfrac{\Big|}{\Big|}}\cr
& & \qquad\qquad\quad
-\;s_\xi^2c_\xi^2
\big[\Cn,\Dn\big]_{\lambda,\eta}
\underbrace{
\left(
\dfrac{-2(1-k^2)\,\Sn(\phi+mt)\,\Sn\left(\tfrac{\Delta m\,t}{2}\right)}
{1-k^2\,\Sn^2\left(\phi+mt\right)\Sn^2\left(\frac{\Delta m\,t}{2}\right)}
\right)}_{\displaystyle
\equiv\mathcal{I}(\phi+mt,\Delta m\,t)}
\vphantom{\dfrac{\frac{X}{X}}{\frac{X}{X}}}
\Bigg]
,
\vphantom{\dfrac{\Big|}{\Big|}}
\label{NambufBsquared}
\end{eqnarray}
where $\Theta_1$ and $\Theta_2$ are, 
respectively, the angles between the pairs of unit vectors
\begin{equation}
\left\{
\begin{bmatrix}
\Cn_\lambda \\ \Sn_\lambda
\end{bmatrix}
,
\begin{bmatrix}
\Cn_\eta \\ \Sn_\eta
\end{bmatrix}
\right\}
,\qquad\mbox{and}\qquad
\left\{
\begin{bmatrix}
k\Sn_\lambda \\ \Dn_\lambda
\end{bmatrix}
,
\begin{bmatrix}
k\Sn_\eta \\ \Dn_\eta
\end{bmatrix}
\right\}
.
\end{equation}
The sign of $\sin\Theta_1$ has been chosen so that
\begin{equation}
\sin\Theta_1 \;\xrightarrow{k\to 0}\; \sin(\eta-\lambda)\;.    
\end{equation}
%
%
%
%
We note that $\cos\Theta_1$ and $\sin\Theta_1$ can take on any value in the range $[-1,1]$, whereas the range of
$\cos\Theta_2$ is limited to
\begin{eqnarray}
1-2k^2 \;\le & \cos\Theta_2 & \le\; 1\;.
\vphantom{\Big|}
\label{Thera1Theta2bounds}
\end{eqnarray}
Since $\Theta_1$ and $\Theta_2$ are both functions of
$\lambda$ and $\eta$, their values are correlated and not all possible values of $\Theta_1$ and $\Theta_2$ in their respective ranges can be simultaneously realized.
To see this, we show in FIG.~\ref{Theta1Theta2Plot} the contour plots of $\cos\Theta_1$, $\sin\Theta_1$, and $\cos\Theta_2$
on the $(\lambda,\eta)$ plane for a single period in each direction.
$k^2$ is taken to be $0.5$ in these plots, but the basic features will remain the same
for any value of $k^2\in(0,1)$.
For instance, the maxima of $\cos\Theta_1$, $\sin\Theta_1$, and $\cos\Theta_2$ will be along the red lines,
and their minima will be along/at the white lines/dots for any value of $k^2$.

Note that $\cos\Theta_1=1$ necessarily implies $\cos\Theta_2=1$, since both functions are at their
maxima of 1 along the $\eta=\lambda$ line.
However, we have $\cos\Theta_2=1$ along the $\eta=-\lambda\pm\pi$ lines as well, which intersect the contours
for all possible values of $\cos\Theta_1$ and $\sin\Theta_1$.
Also, we have $\cos\Theta_1=-1$, $\sin\Theta_1=0$ along the lines $\eta=\lambda\pm\pi$, which intersect the contours for all possible value of $\cos\Theta_2$.
Therefore, any value of $\cos\Theta_1$ and $\sin\Theta_1$ can be realized when $\cos\Theta_2=1$,
and any value of $\cos\Theta_2$ can be realized when $\cos\Theta_1=-1$, $\sin\Theta_1=0$.
However, an arbitrary contour for $\cos\Theta_1$ does not necessarily cross all the contours for $\cos\Theta_2$,
and vice versa, \textit{e.g.} the $\cos\Theta_1=1$ contour, so not all value pairs can be realized.
%
%
%
%
%
%
%
%
%

\begin{figure}[t]
\includegraphics[width=17.5cm]{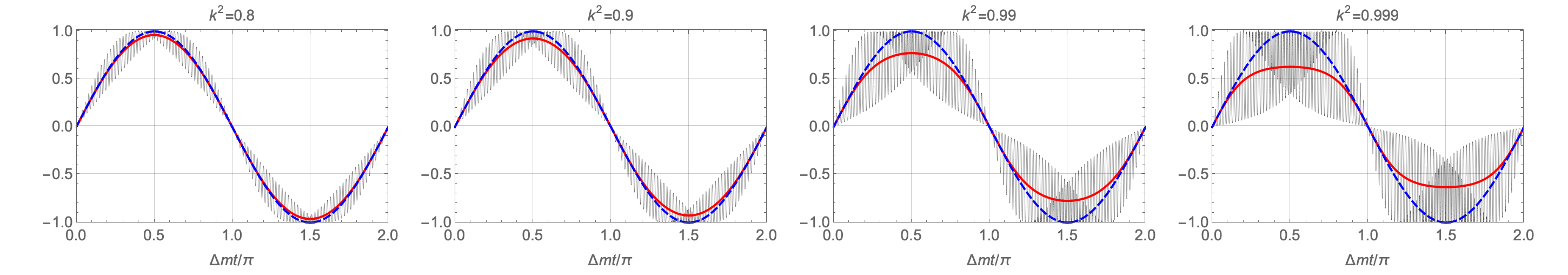}
\caption{Comparison of $\mathcal{G}(mt,\Delta m\,t)$ (gray), $\overline{\mathcal{G}}(\Delta m\,t)$ (red),
and $\sin(\Delta m\,t)$ (blue dashed) for $k^2=0.8, 0.9, 0.99, 0.999$,
for the case $2m/\Delta m=100$.
}
\label{Gplot}
\end{figure}

$|\braket{f}{\overline{B^0}(t)}|^2$
is the same as $|\braket{f}{B^0(t)}|^2$
except the sign of the $A_L A_H$ term is reversed.
Therefore,
\begin{eqnarray}
A(t)
& = &
\dfrac{2A_L A_H}{A_L^2 + A_H^2}
\bigg[
-\Big\{
c_\xi^2\cos\Theta_1 + s_\xi^2\cos\Theta_2
\Big\}
\Big\{
1 - 2\mathcal{F}(\phi+mt,\Delta m\,t)
\Big\}
\vphantom{\Bigg|}\cr
& & \qquad\qquad\quad +\;c_\xi^2\left(c_\xi^2+s_\xi^2 k^2\right)
\sin\Theta_1
\;\mathcal{G}(\phi+mt,\Delta m\,t)
\vphantom{\Bigg|}\cr
& & \qquad\qquad\quad
-\;s_\xi^2 \left(c_\xi^2+s_\xi^2 k^2\right)
\big[\Sn,\Dn\big]_{\lambda,\eta}
\;\mathcal{H}(\phi+mt,\Delta m\,t)
\vphantom{\Bigg|}\cr
& & \qquad\qquad\quad 
+\;s_\xi^2 c_\xi^2
\big[\Cn,\Dn\big]_{\lambda,\eta}
\;\mathcal{I}(\phi+mt,\Delta m\,t)
\bigg]
\;.
\vphantom{\Bigg|}
\end{eqnarray}
Here we use the fact that $m\gg \Delta m$ to replace the
$\mathcal{F}(\phi+mt,\Delta m\,t)$,
$\mathcal{G}(\phi+mt,\Delta m\,t)$,
$\mathcal{H}(\phi+mt,\Delta m\,t)$, and
$\mathcal{I}(\phi+mt,\Delta m\,t)$ functions with those that are averaged
over a period of the first argument.
It is straightforward to see that
\begin{eqnarray}
\overline{\mathcal{H}}(y)
& = & \dfrac{1}{2\pi}\int_{-\pi}^{\pi}dx\,\mathcal{H}(x,y) \;=\; 0\;,\cr
\overline{\mathcal{I}}(y)
& = & \dfrac{1}{2\pi}\int_{-\pi}^{\pi}dx\,\mathcal{I}(x,y) \;\;=\; 0\;.
\end{eqnarray}
Graphs of the function $\overline{\mathcal{G}}(\Delta m\,t)$, where
\begin{equation}
\overline{\mathcal{G}}(y) \;=\; \dfrac{1}{2\pi}\int_{-\pi}^{\pi}dx\,\mathcal{G}(x,y)\;,
\end{equation}
is compared to that of $\mathcal{G}(mt,\Delta m\,t)$ in FIG.~\ref{Gplot}
for several values of $k^2$, where the mass ratio was taken to be $2m/\Delta m=100$.
Using the $k^2$-expansions of the Jacobi elliptic functions given in Appendix~\ref{Jacobi}, we can demonstrate that
\begin{equation}
\overline{\mathcal{G}}(\Delta m\,t)
\;=\;
\sin(\Delta m\,t) + O(k^4)\;.
\end{equation}
The graph of $\sin(\Delta m\,t)$ is also shown in FIG.~\ref{Gplot} and we can see that the
approximation is very good until $k^2$ exceeds $0.9$ or so.

Thus, we arrive at the expression for $A(t)$ in Nambu QM:
\begin{eqnarray}
A(t) & = &
\dfrac{2A_L A_H}{A_L^2+A_H^2}
\bigg[
c_\xi^2\left(c_\xi^2+s_\xi^2 k^2\right)
\sin\Theta_1
\;\overline{\mathcal{G}}(\Delta m\,t)
\cr
& & \qquad\qquad\quad
-\Big\{
c_\xi^2\cos\Theta_1 + s_\xi^2\cos\Theta_2
\Big\}
\Big\{
1 - 2\overline{\mathcal{F}}(\Delta m\,t)
\Big\}
\bigg]
\;.
\vphantom{\Bigg|}
\label{AtExact}
\end{eqnarray}
For $k=0$, this simplifies to
\begin{equation}
A(t;k=0)
\;=\; \dfrac{2A_L A_H}{A_L^2+A_H^2}
\bigg[
c_\xi^4 \sin(\eta-\lambda)\sin(\Delta m\,t)
- 
\Big\{
c_\xi^2\cos(\eta-\lambda) + s_\xi^2
\Big\}
\Big\{
c_\xi^2\cos(\Delta m\,t) + s_\xi^2
\Big\}
\bigg]
\;.
\label{AtNQMk0}
\end{equation}
If we further set $\xi=0$, we recover the canonical QM expression Eq.~\eqref{AtCanonialQM}.
On the other hand, if we set $k=0$, $\xi=\pi/2$, the limit in which oscillations vanish, the expression becomes
\begin{equation}
A\big(t;k=0,\xi=\tfrac{\pi}{2}\big) \;=\; -\dfrac{2A_L A_H}{A_L^2 + A_H^2}\;,
\end{equation}
that is, $A(t)$ will be constant as expected.  
Note that since $A_L$ and $A_H$ can be either positive or negative in Nambu QM, this constant can be
any number in the range
\begin{equation}
-1 \;\le\; -\dfrac{2A_L A_H}{A_L^2 + A_H^2} \;\le\; 1\;.
\end{equation}
In this pseudo-classical limit, the $B^0$ state stays a $B^0$ state, and the
$\overline{B^0}$ state stays a $\overline{B^0}$ state.  The ratio of their decay rates to $\ket{f}$ is
\begin{equation}
\dfrac{|\braket{f}{\overline{B^0}}|^2}{|\braket{f}{B^0}|^2}
\;=\; \bigg(\dfrac{A_L-A_H}{A_L+A_H}\bigg)^2
\;,
\end{equation}
which can be any non-negative number.

\begin{figure}[ht]
\begin{center}
\includegraphics[width=17.5cm]{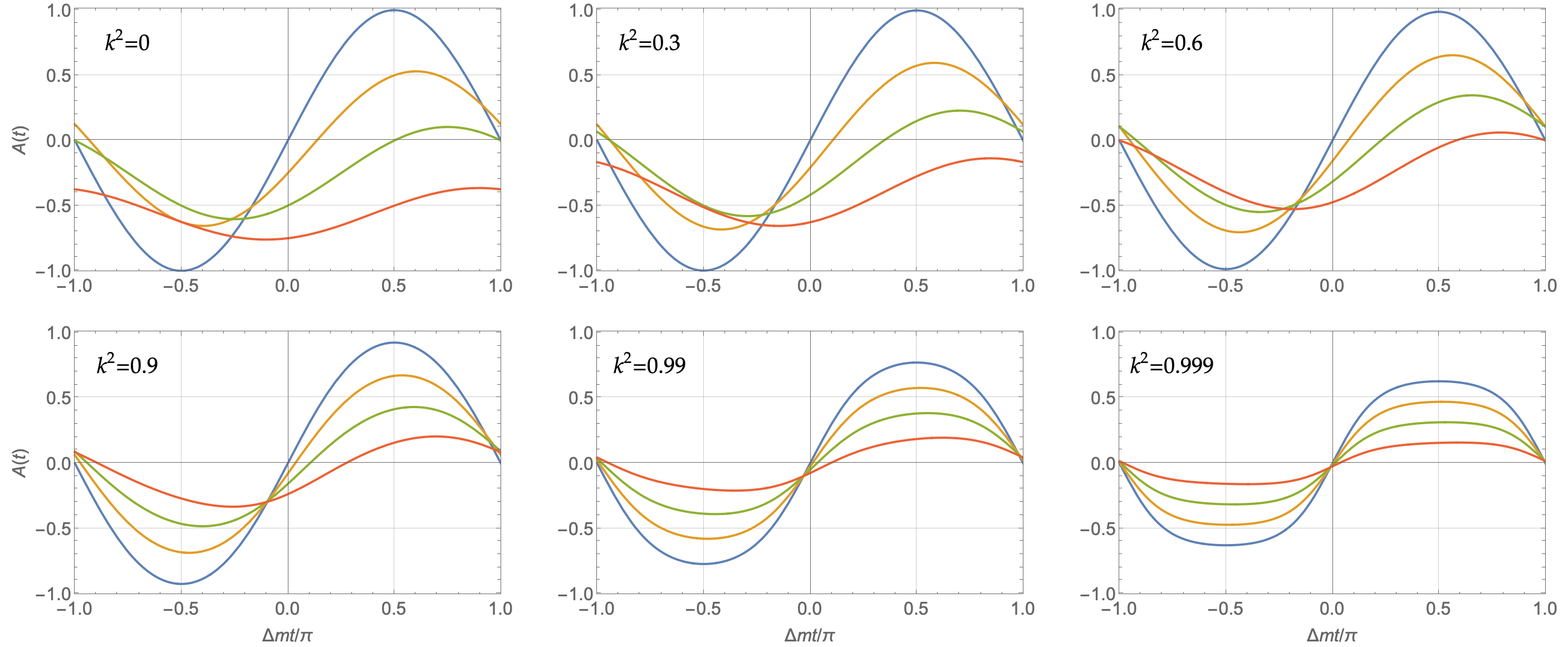}
\end{center}
\caption{Graphs of $A(\Delta m\,t)$
for $2A_L A_H/(A_L^2+A_H^2)=1$, $\lambda=0$, $\eta=\pi/2$.
The four curves in each graph are for
$\xi=0$ (blue), $\xi=\pi/6$ (yellow), $\xi=\pi/4$ (green), and $\xi=\pi/3$ (red). 
The canonical QM case is $k^2=0$ and $\xi=0$, which is the blue curve in the leftmost graph in the upper row.
When $k^2=0$, the graph becomes flat at $A(t)=-1$ in the limit $\xi=\pi/2$.
}
\label{AtPlot}
\end{figure}

In FIG.~\ref{AtPlot}, we illustrate the dependence of $A(t)$ on $\xi$ and $k$.
There, the other parameters are fixed to
\begin{equation}
\dfrac{2A_L A_H}{A_L^2 + A_H^2}\;=\;1\;,\qquad
\lambda \;=\; 0\;,\qquad
\eta \;=\; \dfrac{\pi}{2}\;,
\end{equation}
for which the $k=\xi=0$ case is $A(t)=\sin(\Delta m\,t)$, whereas 
the $k=0$, $\xi=\pi/2$ case is $A(t)=-1$.
The deviation of the curve from sinusoidal is small until $k^2$ exceeds $0.9$ or so.
Increasing $\xi$ toward $\pi/2$ when $k=0$ suppresses the oscillation amplitude and shifts the graph toward the flat no-oscillation limit.

\subsection*{Fit to Belle Data}

Let us now use actual Belle data to constrain the deformation parameters $k$ and $\xi$.
We use the data from \cite{Belle1}.
We first consider the $k=0$ case and look at how the fit of Eq.~\eqref{AtNQMk0} 
to the Belle data constrains the parameter $\xi$.

In interpreting the experimental data, one should take into account the miss-tag factors $w_{B^0}$ and $w_{\overline{B}^0}$ that 
respectively characterizes the probability of incorrectly tagging the $B^0$ as $\overline{B^0}$, and vice versa. 
The probability amplitudes are modified to
\begin{eqnarray}
\big|\braket{f}{B^0(t)}\big|^2 
& \rightarrow & (1-w_{B^0})\big|\braket{f}{B^0(t)}\big|^2+w_{\overline{B}^0}\big|\braket{f}{\overline{B}^0(t)}\big|^2 \;,\cr
\big|\braket{f}{\overline{B^0}(t)}\big|^2 
& \rightarrow & (1-w_{\overline{B}^0})\big|\braket{f}{\overline{B^0}(t)}|^2+w_{B^0}|\braket{f}{B^0(t)}\big|^2\;,
\end{eqnarray}
and $A(t)$ to
\begin{eqnarray}
A(t) & \to & 
\dfrac{\Big[(1-w_{\overline{B}^0})\big|\braket{f}{\overline{B^0}(t)}|^2+w_{B^0}|\braket{f}{B^0(t)}\big|^2\Big]
      -\Big[(1-w_{B^0})\big|\braket{f}{B^0(t)}\big|^2+w_{\overline{B}^0}\big|\braket{f}{\overline{B}^0(t)}\big|^2\Big]}
      {\Big[(1-w_{\overline{B}^0})\big|\braket{f}{\overline{B^0}(t)}|^2+w_{B^0}|\braket{f}{B^0(t)}\big|^2\Big]     
      +\Big[(1-w_{B^0})\big|\braket{f}{B^0(t)}\big|^2+w_{\overline{B}^0}\big|\braket{f}{\overline{B}^0(t)}\big|^2\Big]}
\cr
& & =\; \Delta w + D A(t)\;,\vphantom{\bigg|}
\end{eqnarray}
where $D=1-2\vev{w}$ is the dilution factor,  $\vev{w}=(w_{B^0}+w_{\overline{B^0}})/2$ the average miss-tag probability, and
$\Delta w = w_{B^0}-w_{\overline{B}^0}$.

The dilution factor $D$ simply suppresses the overall amplitude $2A_L A_H/(A_L^2 + A_H^2)$ of $A(t)$.
Since $2A_L A_H/(A_L^2 + A_H^2)$ is already restricted to have magnitude smaller than one, the presence of $D$ does not add any 
new constraints.
The difference in miss-tag probabilities, $\Delta w$, leads to an offset of $A(t)$ in the vertical direction.
This is known to be negligibly small for analyses involving a low number of events, and indeed we find that its preferred value
when included in our fits is 
approximately zero.
Therefore, we set $\Delta w=0$ in the following.

\par
\begin{figure}[ht]
\begin{center}
\includegraphics[width=8cm]{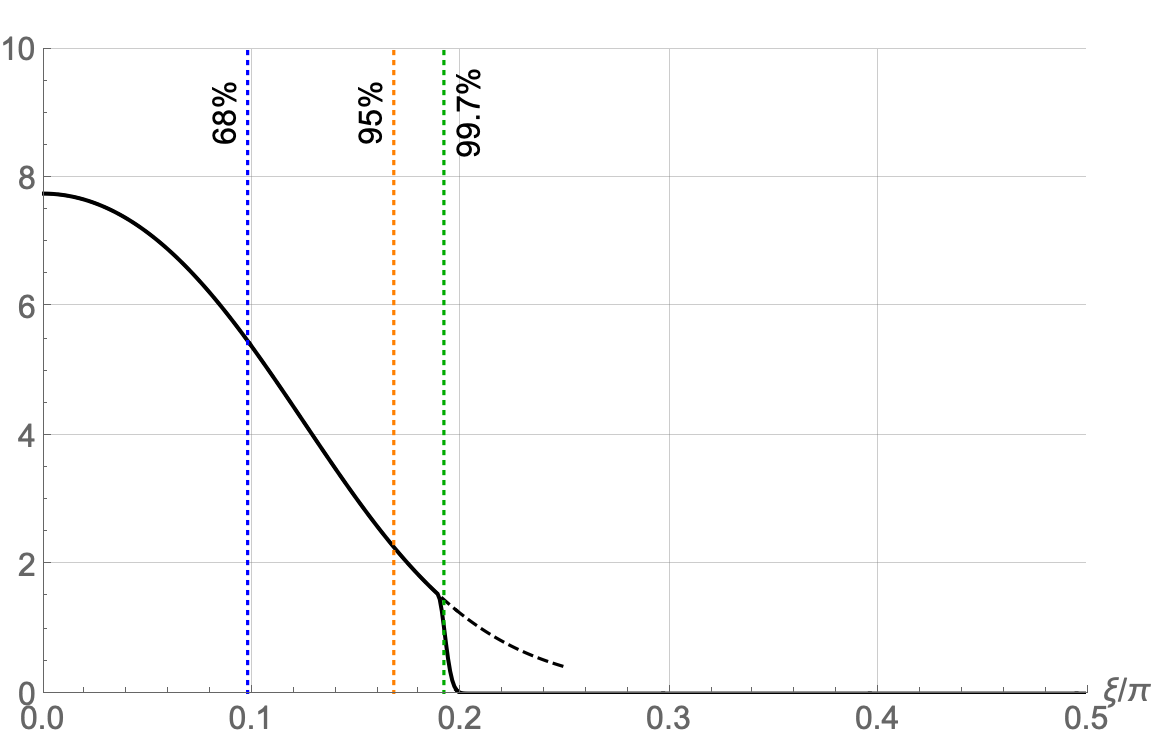}
\end{center}
\caption{The likelihood function for $\xi$ for the $k=0$ case. 
The vertical axis is in units so that the area underneath the curve
is one.  The sudden drop at $\xi/\pi\sim 0.2$ is due to the amplitude $|2A_L A_H/(A_L^2 + A_H^2)| \le 1$ being unable to
compensate for the flattening of the oscillation curve as $\xi$ is increased, \textit{cf.} FIG.~\ref{AtPlot}.
The dashed line indicates how the likelihood function would have looked without the amplitude constraint.
The vertical dotted lines indicate the locations of the $1\sigma$ (68\%), $2\sigma$ (95\%), and $3\sigma$ (99.7\%) bounds.
}
\label{xiLikelihood}
\end{figure}

\begin{figure}[ht]
\begin{center}    
    \includegraphics[width=9cm]{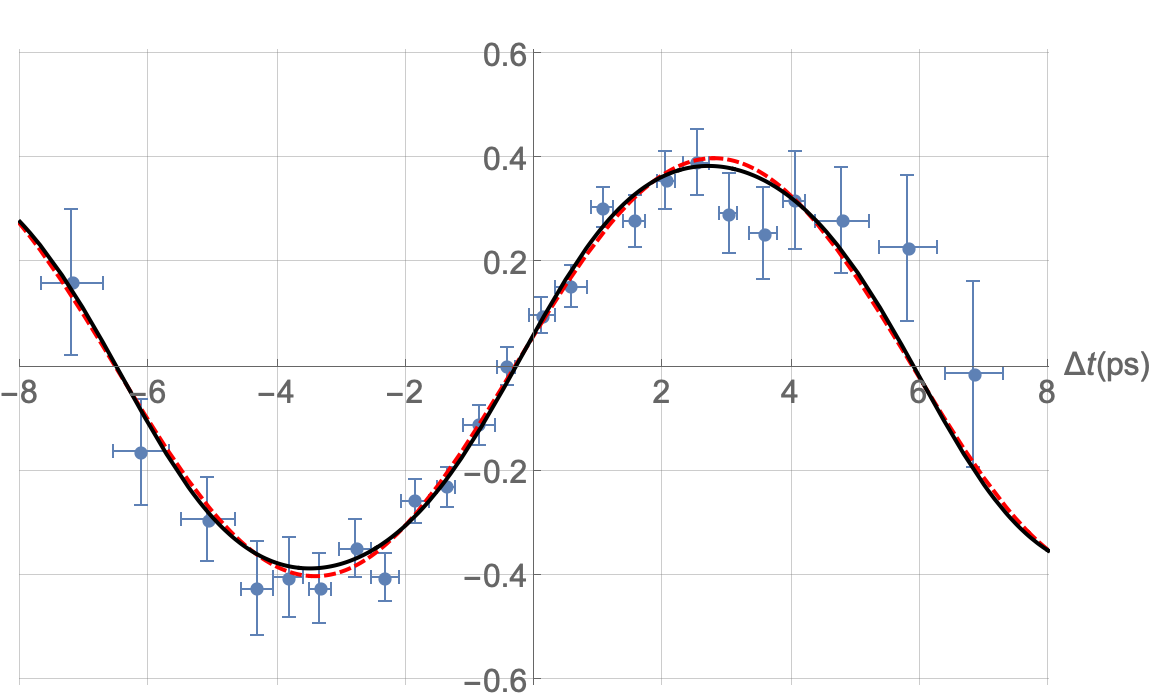}
\end{center}
\caption{Time-dependent CP asymmetry data from Belle \cite{Belle1} fitted with Nambu QM (black solid) and canonical QM (red dashed). For the Nambu QM curve, $k^2=0.957$ and $\xi=0$. The value of the $\chi^2$ per degree of freedom at the best fit point is
$14.32/(24-5)=0.75$ for Nambu QM, and $15.06/(24-2)=0.68$ for canonical QM. 
}
\label{dmcomp}
\end{figure}

 Firstly, we fit Eq.~\eqref{AtNQMk0} to the Belle data allowing 
$2A_L A_H/(A_L^2+A_H^2)$, $(\eta-\lambda)$, and $\xi$ to float.
The mass difference $\Delta m$ is set to the current world average, Eq.~\eqref{PDBDeltam}, from the Review of Particle Properties \cite{ParticleDataGroup:2022pth}.
The minimum of the $\chi^2$ is found to be $15.06$ for $24-3$ degrees of freedom at
\begin{equation}
\dfrac{2A_L A_H}{A_L^2+A_H^2}\;=\; \pm 0.4006\;,\qquad 
\dfrac{\eta-\lambda}{\pi}\;=\; \begin{cases} +0.5496 \\ -0.4504 \end{cases},\qquad 
\xi\;=\; 0\;,
\end{equation}
that is, canonical QM is preferred over a non-zero $\xi$.
The best-fit curve is shown against the data of \cite{Belle1} in FIG.~\ref{dmcomp}
as the canonical QM fit.
The likelihood function for $\xi$ is shown in FIG.~\ref{xiLikelihood}.
It has the characteristic feature that it suddenly drops to zero around $\xi/\pi\sim 0.2$.
This drop is caused by the amplitude $|2A_L A_H/(A_L^2 + A_H^2)| \le 1$ being unable to
compensate for the flattening of the oscillation curve as $\xi$ is increased beyond this point.
This is similar to the atmospheric neutrino analysis in which $\sin^2 2\theta \le 1$ was unable to
compensate for an increase in $\xi$, and the resulting suppression of $c_\xi^2$.
From this likelihood curve, we find the upper bounds on $\xi$ to be 
\begin{equation}
\dfrac{\xi}{\pi} \;<\;
0.098\;\;(1\sigma,\;68\%),\qquad
0.168\;\;(2\sigma,\;95\%),\qquad
0.192\;\;(3\sigma,\;99.7\%)\;.
\label{mzeroBelleBounds}
\end{equation}
Note that these are weaker than the atmospheric neutrino bounds given in Eq.~\eqref{mzeroAtmosNuBounds}.

\begin{figure}[ht]
\begin{center}
\includegraphics[width=7.5cm]{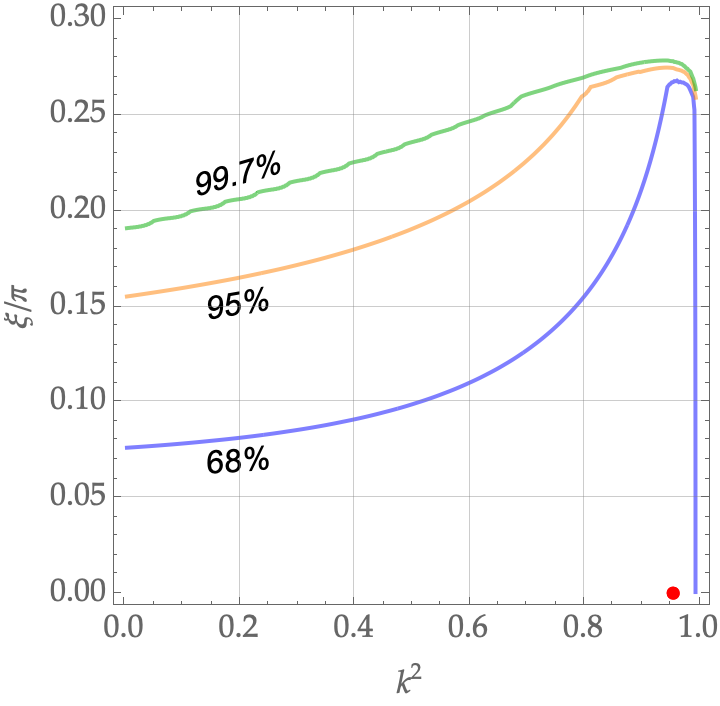}
\hspace{1cm}
\includegraphics[width=7.5cm]{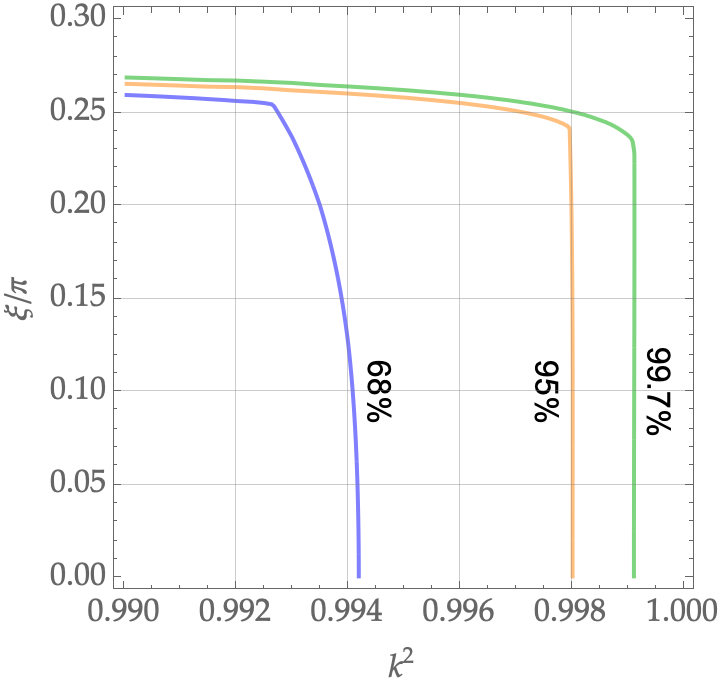}
\end{center}
\caption{$k^2$-$\xi$ contour plot showing the $1\sigma$(68\%), $2\sigma$(95\%) and $3\sigma$(99.7\%) likelihood contours for the
Belle fit.   The red dot denotes the point which minimizes the $\chi^2$.
The right figure is a blowup of the region $0.99<k^2<1$.}
\label{BelleContour}
\end{figure}

Next, we fit Eq.~\eqref{AtExact} to the Belle data to constrain both $\xi$ and $k$.
The fit parameters are $2A_L A_H/(A_L^2+A_H^2)$, $\lambda$, $\eta$, $k^2$, and $\xi$.
The mass difference $\Delta m$ is set to Eq.~\eqref{PDBDeltam}.
The minimum of the $\chi^2$ is found to be $14.32$ for $24-5=19$ degrees of freedom when
\begin{equation}
\dfrac{2A_L A_H}{A_L^2+A_H^2}\;=\; \pm 0.4403\;,\qquad 
\dfrac{\Theta_1(\lambda,\eta)}{\pi}\;=\; 
\begin{cases}
+0.5469\\
-0.4531
\end{cases},\qquad
k^2 \;=\; 0.957\;,\qquad
\xi \;=\; 0\;.
\label{BelleBestFitParams}
\end{equation}
$\Theta_2(\lambda,\eta)$ is unconstrained due to the best fit value of $\xi$ being zero,
leading to degeneracies on the $(\lambda,\eta)$ plane. 
The best fit curve is shown in FIG.~\ref{dmcomp} against the data and the canonical QM best fit.
The likelihood contours on the $k^2$-$\xi$ plane are shown in FIG.~\ref{BelleContour}.
Note that the best fit point is at $\xi=0$ and $k^2=0.957$, exceptionally close to the maximum allowed value of $k^2$. 
This is due to the fact that the data prefers an oscillation which slightly deviates from the sinusoidal, 
as can be seen in FIG.~\ref{dmcomp},
while Jacobi's elliptical functions show such deviations in a noticeable way only as $k^2$ approaches 1.
This leads to $k^2$ being only very weakly constrained.
The $3\sigma$ contour on FIG.~\ref{BelleContour} traces a precipice in the likelihood function along which it plummets to zero,
just as in the $k=0$ case, \textit{cf.} FIG.~\ref{xiLikelihood}.
This is again due to the constraint $|2A_L A_H/(A_L^2+A_H^2)|\le 1$.

\section{Summary and Discussion}\label{Summary}

In this paper, we have reviewed how Nambu QM \cite{Minic:2002pd} extends canonical QM via its geometrical formulation \cite{Kibble:1978tm,Ashtekar:1997ud,Brody:1999cw,Anandan:1990fq,Cirelli:1999in,Cirelli:2003}.
All states are expressed as superpositions of energy eigenstates with
coefficients that have magnitudes and ``phases.'' 
The extension involves generalizing the concept of the ``phase'' to a map from $S^1$ (a circle) to $S^2$ (a sphere).  
The $2N$ dimensional K\"ahler manifold of geometric QM for $N$ energy eigenstates is generalized to a $3N$ dimensional manifold.
Canonical QM corresponds to the ``phase'' being a map from $S^1$ to the equator of $S^2$.

Deviations of the map from the equator is described by two deformation parameters $\xi$ and $k$.
The trajectory of the map is assumed to follow the path of the angular momentum of an asymmetric top,
the motion of which can be described by the triple Nambu bracket \cite{Nambu:1973qe} with two conserved quantities, one of which maintains
the ``phase'' vector on the surface of $S^2$.
The dependence of the 3D ``phase'' vector on the phase parameter is described by Jacobi's
elliptical functions $\cn(u,k)$, $\sn(u,k)$, and $\dn(u,k)$, where $k$ is the eccentricity of the ellipse 
from which these function are defined. 

Under this deformation, the coefficients of the energy eigenstates are no longer elements of a field or division algebra
since the addition of coefficients is ill defined.  
The inner product defined for the extension is also not invariant
under generic phase shifts of the coefficients.  Therefore, the projective vector space structure of the state space
of canonical QM is lost.  
Since the coefficients do not add, there is no path-integral formulation of the Nambu extension in which
such a coefficient is associated with each path, and the paths interfere with each other via superposition.
Nevertheless, the energy eigenstates with those coefficients do interfere allowing for oscillation phenomena, which 
deform those of canonical QM.

We have used atmospheric neutrino \cite{deSalas:2020pgw} and Belle \cite{Belle1} data to constrain the deformation parameters $\xi$ and $k$, as
shown in FIGS.~\ref{AtmosBounds} and \ref{BelleContour}.
Of the two deformation parameters $\xi$ is the better constrained, the $3\sigma$ upper bounds for the $k=0$ case being
$\sim 0.07\pi$ and $\sim 0.2\pi$ for atmospheric neutrinos and Belle, respectively,
\textit{cf.} Eq.~\eqref{mzeroAtmosNuBounds} and \eqref{mzeroBelleBounds}.
This is due to the parameter $\xi$ suppressing oscillations as it is increased from 0 to $\pi/2$, while
the overall amplitude ($\sin^2 2\theta$ in the case of atmospheric neutrinos and $2A_L A_H/(A_L^2+A_H^2)$ in the case of
$B^0$-$\overline{B^0}$ oscillations, which are both bounded by 1) is unable to compensate for it.
The somewhat surprising result is that the bound from atmospheric neutrinos is stronger than the Belle bound.

The bound on $k^2$ in contrast is very weak.
This is due to $k^2$ causing significant deviations of the oscillation functions from the sinusoidal only when $k^2$ is very close to one.
Indeed, the best fit value to the Belle data was $k^2=0.957$, Eq.~\eqref{BelleBestFitParams}, 
due to the data preferring a slight deviation of the oscillation curve from the sinusoidal.  See FIG.~\ref{dmcomp}.
However, this was based on a simplified oscillation formula which resulted due to the sum of the eigenfrequencies of the interfering energy eigenstates being much larger that the difference, allowing us to average out the fast oscillations.
For situations in which the sum and difference of the frequencies are much closer to each other, there will be significant deviations from 
sinusoidal oscillations at the frequency difference, and we expect the bound on $k^2$ to be much stronger.
Furthermore, the $k=0$ case is preferred theoretically, since it will allow for phase changes of the energy eigenstates without
affecting physical probabilities, justifying the somewhat arbitrary phase choices in Eqs.~\eqref{FlavorEigenstates} and \eqref{B0B0barStates}.

In addition to the loss of the projective vector space structure of the state space, Nambu QM has other shortcomings as well.
First, unstable states cannot be made to decay by giving them complex energy eigenvalues.
This is due to the doubly periodic nature of the Jacobi elliptic functions which are periodic in the imaginary direction as well as the real.
Thus, we must let the amplitudes of the coefficients decay exponentially by hand.
Second, since the Nambu extension relies on a fixed energy-eigenstate basis, it is not clear how it should be applied to
situations in which the energy eigenstates and their respective eigenvalues are themselves time-dependent.
For instance, when considering matter effects on neutrino oscillations, the eigenstates of the neutrino Hamiltonian 
change depending on the background electron density.
This is the reason why we did not analyze neutrino oscillations directly in this paper.

Despite these drawbacks, Nambu QM has many interesting features beyond the deformation of oscillations that can be
constrained by experiment.
First, when $k=0$ but $\xi\neq 0$, the inner product becomes invariant under a global phase shift of all states by a common phase.
Thus, if we gauge the theory it will become a $U(1)$ gauge theory in which all states carry a common $U(1)$ charge.
This may provide a new way to quantize $U(1)$ charges without the need to embed the theory into a non-Abelian gauge theory.

Second, the parameter choice $\xi=\pi/2$, $k=0$ leads to a model which is pseudo-classical, \textit{i.e.} all oscillations vanish
in this limit (though interferences do not).  Thus Nambu QM provides a framework in which one can interpolate between quantum and classical-like behavior continuously by tuning the deformation parameter $\xi$.
This property may allow us to study how the Leggett-Garg inequalities \cite{Leggett:1985zz,Emary:2014A,Emary:2014B,Formaggio:2016cuh,Fu:2017hky}
set in as pseudo-classicality is approached.

We note that the loss of addition of the ``numbers'' that constitute the state coefficients 
may be what is necessary to render a quantum-like theory classical (or pseudo-classical).
Indeed, in \cite{Chang:2013rya} it has been shown that in the $q\to 1$ limit of $\mathbb{F}_q$QM \cite{Chang:2012eh,Chang:2012gg,Chang:2012we,Chang:2019kcp},
which is a quantum mechanical model constructed over the finite Galois field $\mathbb{F}_q$, 
the theory behaves ``classically,'' while $\mathbb{F}_q$ becomes the so-called ``field with one element'' 
$\mathbb{F}_1$, or $\mathbb{F}_{\text{un}}$
\cite{Tits:1957}, for which the elements multiply but do not add.

Third, it may be possible to construct a model with Sorkin's triple-path interference \cite{Sorkin:1994dt,Huber:2021xpx,Berglund:2023vrm} using the Nambu approach.
In both canonical and Nambu QM, interference between different energy eigenstates is pairwise due to the Born Rule.
They result from the cross terms that appear in the squares of the symmetric and anti-symmetric parts of the inner product, namely:
\begin{equation}
(\vec{\Phi}_n\cdot\vec{\Psi}_n)(\vec{\Phi}_m\cdot\vec{\Psi}_m)\;,\quad\mbox{and}\quad
(\vec{\Phi}_n\times\vec{\Psi}_n)\cdot(\vec{\Phi}_m\times\vec{\Psi}_m)\;,\qquad
n\neq m\;.
\end{equation}
However, in the Nambu extension $(\vec{\Phi}_n\times\vec{\Psi}_n)$ is a 3D vector, allowing us to construct the 3D volume
\begin{equation}
\big(\vec{\Phi}_n\times\vec{\Psi}_n\big)\cdot
\Big[
\big(\vec{\Phi}_m\times\vec{\Psi}_m\big)\times
\big(\vec{\Phi}_k\times\vec{\Psi}_k\big)
\Big]\;,\qquad
n\neq m\neq k\neq n\;,
\end{equation}
which could be interpreted as the interference of the $n$th, $m$th, and $k$th energy eigenstates.
If such a term can be incorporated into a unitary model, we would have an extension of canonical QM with triple-path
interference.  
Note that the existence of such a model is crucial in experimentally probing the validity of the Born rule \cite{Sinha418,Huber:2021xpx}
as an alternative theory against which to compare canonical QM.

Furthermore, though we only consider an extension of the ``phase'' to a map from $S^1$ to $S^2$,
the Nambu bracket \cite{Nambu:1973qe} exists for any number of dimensions and it can be used to extend the ``phase'' to
a map from $S^1$ to $S^{n-1}$, $n=4,5,\cdots$.  In those higher dimensional extensions, the phase vector will be $n$-dimensional,
and $n$-dimensional hypervolumes can be constructed, which could correspond to $n$-path interferences.
Therefore, the possibility exists for a ladder of models, the $n$-dimensional case possessing Sorkin's multi-path interference up to the $n$th order. 
However, if one insists on the top-like behavior of the phase vector, then one may need to jump to 7D \cite{Grabowski:1992dt,Fairlie:1997sj,Ueno:1998en,Silagadze_2002}
in which the model will have connections to octonions \cite{Gursey:1996mj}.

If we do not restrict our attention to models in which the phase vector is top-like, then even in the 3D case
we can use ``phase'' maps from $S^1$ to $S^2$ that are different from what was used in this paper.
For instance, we could let the phase vector of each energy eigenstate follow
different great circles on $S^2$.  
Each choice of ``phase'' map will lead to a different phenomenology from what was discussed.

The possibility of multi-path interferences also suggests the relevance of Nambu QM to the quantization of gravity.
Indeed, in the new approach to quantum gravity dubbed ``gravitization of quantum theory'' \cite{Freidel:2014qna,Berglund:2022skk}, 
the phenomena of triple and higher order interference are naturally expected to occur \cite{Berglund:2023vrm}.
The reason for this stems from the fact that ``gravitization of quantum theory'' by definition implies a dynamical geometry of quantum theory, and thus the generalization of the Born rule,
which is tied to the canonical geometry of complex projective spaces. This canonical geometry is maximally symmetric and via the Born rule it precludes any intrinsic triple and higher order interference. However, given the non-linear and non-polynomial nature of gravitational interaction, it is expected that ``gravitization of quantum theory'' will lead to triple and higher interference of arbitrary order.

Other avenues worth exploring include the quantum brachistochrone problem, which is usually stated as follows:
Find the Hamiltonian that takes one from a given initial state to a given final state in the least amount of time, under the constraint that the difference between the largest and the smallest eigenvalues of the Hamiltonian is fixed \cite{Carlini:2005usu,Bender:2006fe,Brody_2006,Brody_2007}. 
Since we have not introduced a Hamiltonian operator in Nambu QM, and also since we can expect transition speeds to depend on the
sums of energy eigenvalues as well as the differences when $k\neq 0$, the problem must be restated somewhat to:
Find the energy spectrum that takes one from a given initial state to a given final state in the least amount of time,
under the constraint that the largest and the smallest energy eigenvalues are fixed.
In particular, how does the answer depend on $k$ and $\xi$ and what happens in the pseudo-classical limit $k=0$, $\xi\to\pi/2$?
Answering this will provide us with a deeper understanding of the dynamics of Nambu QM, and perhaps
illuminate what happens in the classical limit of a quantum theory. 

Classical Nambu theory fits into the general Hamiltonian framework of the principle of least action $\delta S=0$, where the action of classical Nambu dynamics is
the integral of a 2-form involving 2 Hamiltonians \cite{Takhtajan:1993vr} (as opposed to the integral of a 1-form involving one Hamiltonian, valid in canonical classical theory).
In general, one has integrals over $n$-forms, with $n$-Hamiltonians. The corresponding quantum theory can be formulated as an $\hbar$ deformation of the classical action principle, by invoking the Schwinger variational principle 
$\delta S \cdot = i \hbar \delta \cdot$, where in the usual case 
one acts ($\delta$) on a complex wavefunction, or alternatively the real and imaginary parts of the wave function, respectively. The real and imaginary parts of the wavefunctions are dual to each other in this formulation that gives the canonical Schr\"{o}dinger evolution. In quantum Nambu theory, the action $\delta$ is on one of the components of a three-component (vector-like) wavefunction, where the dual to each component is given by the cyclic product of the other two components. Obviously this can be generalized to $n$-components. In this case the variation of the action $\delta S$, in general, involves $n$ Hamiltonians of the Nambu dynamics. This general formulation leads in principle to a new view of how probabilities are calculated in Nambu quantum theory. From this point of view, Nambu quantum theory fits into a more general understanding of quantum mechanics as quantum measure theory \cite{Sorkin:1994dt}, with general probabilities related to triple \cite{Huber:2021xpx} and higher order interference phenomena
\cite{Berglund:2023vrm}.

In this context another curious question is what happens when we discretize the ``phase'' space variables of Nambu QM. This question is worth exploring especially in connection to Spekkens' toy model~\cite{Spekkens:2007} and its generalization, dubbed Quasi-quantization~\cite{Spekkens2016}, in which one considers classical physical systems and imposes a limitation on the amount of knowledge that can be possessed by an observer about those systems. The theory derived from those assumptions is able to qualitatively reproduce various phenomena that are ordinarily considered characteristic of quantum theory. For instance, the version of such theory in which the phase space variables $p$ and $q$ are restricted to take on values in $\mathbb{F}_2$ can reproduce a large subset of phenomena associated with quantum spins. Among the phenomena reproduced therein, the ones that are of most interest to us are coherent superposition and interference, owing to our goal of modeling triple-path interference. As such, our future work will explore Quasi-quantization of theories with three ``phase'' space variables in general and their discrete version in particular. However, we need to have a better understanding of the third ``phase'' space variable before attempting this study. This is because the analog of epistemic restriction in quantum theory is Heisenberg's uncertainty principle and trying to define its three-variable version deserves careful consideration.

There remain numerous avenues to be pursued in further development and understanding of Nambu QM. 
We intend to address these possibilities in our forthcoming publications.

\section{Acknowledgements}
The authors would like to thank L. Piilonen for 
helpful communications pertaining to the $B^0$-$\overline{B^0}$ oscillation measurements at Belle. 
Discussions with C. H. Tze are also gratefully acknowledged. 
D. Minic and T. Takeuchi thank the Julian Schwinger Foundation and the U.S. Department of Energy (DE-SC0020262) for support.

\appendix
\section{Classical Nambu Dynamics}
Here we review some properties of the Poisson and Nambu brackets.

\subsection{Poisson Bracket}

If we denote
\begin{equation}
q_1 \;=\; q\;,\qquad q_2 \;=\; p\;,
\end{equation}
then the Poisson bracket can be written as
\begin{equation}
\{A,B\} 
\;=\; \dfrac{\partial A}{\partial q}\,\dfrac{\partial B}{\partial p}
     -\dfrac{\partial A}{\partial p}\,\dfrac{\partial B}{\partial q}
\;=\; \varepsilon_{ij}\,\dfrac{\partial A}{\partial q_i}\,\dfrac{\partial B}{\partial q_j}\;,
\end{equation}
where $\varepsilon_{ij}$ is the totally antisymmetric tensor with $\varepsilon_{12}=1$\;.
The Poisson bracket has the following properties:
\begin{enumerate}
\item Anti-symmetry:
\begin{equation}
\{A_1,A_2\} \;=\; -\{A_2,A_1\}\;.
\end{equation}

\item Leibniz Rule (the operator $\{A_1,*\}$ is a derivative acting on $*$) :
\begin{equation}
\{A_1,A_2A_3\} \;=\; \{A_1,A_2\}A_3 + A_2\{A_1,A_3\}\;.
\end{equation}

\item Fundamental Identity: 
\begin{equation}
 \{A_1,\{A_2,A_3\}\}
+\{A_2,\{A_3,A_1\}\}
+\{A_3,\{A_1,A_2\}\}
\;=\; 0\;,
\end{equation}
which can also be written as
\begin{equation}
\{A_1,\{A_2,A_3\}\}
\;=\;
 \{\{A_1,A_2\},A_3\}
+\{A_2,\{A_1,A_3\}\}
\;,
\end{equation}
which is the Leibniz rule of the derivative $\{A_1,*\}$ acting on $\{A_2,A_3\}$.

\end{enumerate}
The equation of motion of observable $F$ is
\begin{equation}
\dfrac{dF}{dt} \;=\; -\{H,F\} \;=\; \{F,H\}\;,
\end{equation}
where $H$ is the Hamiltonian.
The operator $-\{H,*\}=\{*,H\}$ is a derivative operator acting on $*$.
For example, if
\begin{equation}
H 
\;=\; \dfrac{\omega}{2}\left(q_1^2 + q_2^2\right)
\;,
\end{equation}
then
\begin{eqnarray}
\dot{q}_1 & = & \{q_1,H\}
\;=\; \omega q_2 \;,
\cr
\dot{q}_2 & = & \{q_2,H\}
\;=\; -\omega q_1 \;,
\end{eqnarray}
which are the usual equations of motion of the harmonic oscillator.

\subsection{Nambu Bracket}

Consider three classical variables which we write $q_1$, $q_2$, and $q_3$.
Define the order 3 Nambu bracket as
\begin{equation}
\{A,B,C\} \;=\; \varepsilon_{ijk}
\,\dfrac{\partial A}{\partial q_{i}}
\,\dfrac{\partial B}{\partial q_{j}}
\,\dfrac{\partial C}{\partial q_{k}}
\;,
\end{equation}
where $\varepsilon_{ijk}$ is the totally antisymmetric tensor with $\varepsilon_{123}=1$.
This object has the following properties:
\begin{enumerate}
\item Skew-symmetry:
\begin{equation}
\{A_1,A_2,A_3\} \;=\; (-1)^{\epsilon(p)}\{A_{p(1)},A_{p(2)},A_{p(3)}\}\;,
\end{equation}
where $\epsilon(p)$ is the signature of the permutation $p$.

\item Leibniz Rule (the operator $\{A_1,A_2,*\}$ is a derivative acting on $*$) :
\begin{equation}
\{A_1,A_2,A_3A_4\} \;=\; \{A_1,A_2,A_3\}A_4 + A_3\{A_1,A_2,A_4\}\;.
\end{equation}

\item Fundamental Identity: 
\begin{eqnarray}
\lefteqn{\{A_1,A_2,\{A_3,A_4,A_5\}\}}
\cr
& = & \{\{A_1,A_2,A_3\},A_4,A_5\}
+\{A_3,\{A_1,A_2,A_4\},A_5\}
+\{A_3,A_4,\{A_1,A_2,A_5\}\}
\;.
\qquad\qquad
\end{eqnarray}
This is the Leibniz rule for the operator $\{A_1,A_2,*\}$ acting on $\{A_3,A_4,A_5\}$.

\end{enumerate}
These are analogues of the properties of the Poisson bracket listed above.

The equation of motion for Nambu dynamics is
\begin{equation}
\dfrac{dF}{dt} \;=\; -\{H_1,H_2,F\} \;.
\end{equation}
with two Hamiltonians $H_1$ and $H_2$.
The operator $\{H_1,H_2,*\}$ is a derivative operator acting on $*$.
Due to the skew-symmetry of the Nambu bracket, it is clear that
\begin{eqnarray}
\dot{H}_1 & = & -\{H_1,H_1,H_2\} \;=\; 0\;,\cr
\dot{H}_2 & = & -\{H_1,H_2,H_2\} \;=\; 0\;,
\end{eqnarray}
that is, both $H_1$ and $H_2$ are conserved. The flow generated by ternary Nambu brackets is volume-preserving, generalizing the area-preserving flow of Poisson brackets.

\subsection{Asymmetric Top}\label{Asymtop}

As an example, let
\begin{eqnarray}
H_1 & = & \dfrac{1}{2}\biggl[\dfrac{1}{I_1}(q_1)^2+\dfrac{1}{I_2}(q_2)^2+\dfrac{1}{I_3}(q_3)^2\biggr] \;,
\vphantom{\Bigg|}\cr
H_2 & = & \dfrac{1}{2}\Bigl[(q_1)^2+(q_2)^2+(q_3)^2\Bigr] \;,
\vphantom{\Bigg|}
\label{H1H2}
\end{eqnarray}
where $q_i$ $(i=1,2,3)$ are the dynamical variables
and $I_i$ $(i=1,2,3)$ are constants.
Without loss of generality we can assume
$I_1 \le I_2 < I_3$.
Then, the equations of motion are
\begin{eqnarray}
\dot{q}_1 & = & -\left\{q_1,H_1,H_2\right\}
\;=\; \left(\dfrac{1}{I_3}-\dfrac{1}{I_2}\right)q_2\, q_3\;,
\vphantom{\Bigg|}
\label{AsymmetricTopEquation}
\end{eqnarray}
and cyclic permutations. 
These can be identified with the equations of motion of 
the angular momentum $\vec{L}=(q_1,q_2,q_3)$ of an asymmetric top with 
moments of inertia $I_1\le I_2 < I_3$, energy $H_1$, 
and total angular momentum squared $2H_2$, in the frame fixed to the top.
The $I_1 = I_2$ case is the symmetric top limit.

Since both $H_1$ and $H_2$ are conserved, 
the ellipsoid $H_1=\text{constant}$ and the sphere $H_2=\text{constant}$ must
have an intersection in the 3D $(q_1,q_2,q_3)$ space for a solution to exist.
This requires the shortest axis of the ellipsoid to be shorter than
the radius of the sphere while the longest axis of the ellipsoid is longer than the radius of the sphere.
We assume
\begin{equation}
I_1 \le I_2 < \dfrac{H_2}{H_1} < I_3
\label{conditions}
\end{equation}
in which case the two intersections are respectively in the northern and southern hemispheres,
and in the limit $I_1=I_2\to H_2/H_1$ the intersections come together at the equator. 
When this condition is satisfied, 
the solution to the equation of motion is given by the Jacobi elliptic functions, namely:
\begin{eqnarray}
q_1(t) & = & \phantom{-}N_1\;\cn(\Omega t,k)\;, \vphantom{\big|}\cr
q_2(t) & = & -N_2\;\sn(\Omega t,k)\;, \vphantom{\big|} \cr
q_3(t) & = & -N_3\;\dn(\Omega t,k)\;, \vphantom{\big|}
\label{ATopSolution}
\end{eqnarray}
where $N_1$, $N_2$, and $N_3$ are all positive, and
\begin{eqnarray}
\Omega^2 & = & 
\dfrac{2(I_3-I_2)(H_2-I_1 H_1)}{I_1 I_2 I_3}
\vphantom{\Bigg|}\cr
& = & 
2H_1(I_3-I_2)\bigg[\dfrac{(H_2/H_1-I_2)+(I_2-I_1)}{I_1 I_2 I_3}\bigg]
\;,\vphantom{\Bigg|}\cr
k^2 & = & \dfrac{(I_2-I_1)(I_3 H_1-H_2)}{(H_2-I_1 H_1)(I_3-I_2)}
\vphantom{\Bigg|}\cr
& = & \bigg[
\dfrac{(I_2-I_1)}{(I_2-I_1)+(H_2/H_1-I_2)}
\bigg]
\bigg[
\dfrac{(I_3-H_2/H_1)}{(I_3-H_2/H_1)+(H_2/H_1-I_2)}
\bigg]
\;.
\vphantom{\Bigg|}
\end{eqnarray}
This solution traces the intersection 
of the sphere and ellipsoid in the southern hemisphere.
Given Eq.~\eqref{conditions}, we note
that $0\le k^2 <1$ where $k^2=0$ if and only if $I_1=I_2$, while $k^2\to 1$ as $I_2\to H_2/H_1$, provided $I_1\neq I_2$.

Due to the non-linearity of Eq.~\eqref{AsymmetricTopEquation}, the 
normalization constants $N_1, N_2,$ and $N_3$ are also fixed:
\begin{equation}
N_1\,=\,\sqrt{\dfrac{2I_1(I_3 H_1 - H_2)}{(I_3 - I_1)}}\;,\qquad
N_2\,=\,\sqrt{\dfrac{2I_2(I_3 H_1 - H_2)}{(I_3 - I_2)}}\;,\qquad
N_3\,=\,\sqrt{\dfrac{2I_3(H_2 - I_1 H_1)}{(I_3 - I_1)}}\;.
\end{equation}
The deformation parameter $\xi$ introduced in Eq.~\eqref{U1map2S2} is given by
\begin{eqnarray}
\tan^2\xi
\,=\, \left(\frac{N_3}{N_1}\right)^2
& = & \frac{I_3(H_2/H_1 - I_1)}{I_1(I_3 - H_2/H_1)}
\vphantom{\Bigg|}\cr
& = & 
\bigg[
\dfrac{I_3}{I_3-H_2/H_1}
\bigg]
\bigg[
\dfrac{(H_2/H_1-I_2)-(I_2-I_1)}{I_1}
\bigg]
\;.\vphantom{\Bigg|}
\end{eqnarray}
We note that $\tan\xi \to 0$, \textit{i.e.} the trajectory collapses to the equator when $I_1 = I_2 = H_2/H_1$.
In the same limit, we have $\Omega^2 = 0$.
This is because $q_3=0$ on the equator, and the equations of motion reduce to $\dot{q}_1 = \dot{q}_2 = 0$.
The value of $k^2$ in this limit depends on how the limit is taken.
However, since $\Omega=0$ and $\vec{L}$ is fixed, the limit value of $k^2$ is irrelevant for the top.
In the other extreme,
$\tan\xi\to\infty$ as $I_1\to 0$ (the ellipsoid is flattened into the $q_2$-$q_3$ plane) or $I_3\to H_2/H_1$ (the ellipsoid touches the sphere at the poles).

\section{Properties of the Jacobi Elliptical Functions}\label{Jacobi}

Here we will list some properties of the elliptical functions relevant to this work.
The second argument $k$ of the elliptical functions will be suppressed in the following to simplify the expressions, e.g. $\sn(u,k)\to \sn\,u$.

\begin{itemize}
\item Analogs of the Pythagorean Theorem:
\begin{equation}
\sn^2 u + \cn^2 u \;=\; 1\;,\qquad
k^2 \sn^2 u + \dn^2 u \;=\; 1\;.
\label{PythagoreanAnalog}
\end{equation}

\item Derivatives:
\begin{eqnarray}
(\sn\,u)' & = & \phantom{-}\cn \,u \;\dn \,u\;,\cr
(\cn\,u)' & = & -\sn \,u \;\dn \,u\;,\cr
(\dn\,u)' & = & -k^2\,\sn \,u \;\cn \,u\;.
\end{eqnarray}

\item Power series expansion in $u$:
\begin{eqnarray}
\sn\,u & = & u -(1+k^2)\dfrac{u^3}{3!} + (1+14k^2+k^4)\dfrac{u^5}{5!} + \cdots \cr
\cn\,u & = & 1 - \dfrac{u^2}{2!} + (1+4k^2)\dfrac{u^4}{4!} - \cdots \cr
\dn\,u & = & 1 - k^2\dfrac{u^2}{2!} + (4k^2+k^4)\dfrac{u^4}{4!} -\cdots
\end{eqnarray}

\item Addition and subtraction theorems:
\begin{eqnarray}
\sn(u+v) 
& = & \dfrac{\sn\,u\,\cn\,v\,\dn\,v + \sn\,v\,\cn\,u\,\dn\,u}{1-k^2\,\sn^2 u\,\sn^2 v}
\;,
\vphantom{\Bigg|}\cr
\cn(u+v)
& = & \dfrac{\cn\,u\,\cn\,v - \sn\,u\,\sn\,v\,\dn\,u\,\dn\,v}{1-k^2\,\sn^2 u\,\sn^2 v}
\;,
\vphantom{\Bigg|}\cr
\dn(u+v)
& = & \dfrac{\dn\,u\,\dn\,v - k^2\,\sn\,u\,\sn\,v\,\cn\,u\,\cn\,v}{1-k^2\,\sn^2 u\,\sn^2 v}
\;.
\vphantom{\Bigg|}
\label{AdditionTheorem}
\end{eqnarray}
Subtraction theorems can be derived by using 
$\sn(-u)=-\sn(u)$, $\cn(-u)=\cn(u)$, and $\dn(-u)=\dn(u)$.
In particular, note that
\begin{eqnarray}
\sn(u+v)\,\sn(u-v)
& = &
\dfrac{\sn^2 u - \sn^2 v}{1-k^2\,\sn^2 u\,\sn^2 v}
\;,
\vphantom{\Bigg|}\cr
\cn(u+v)\,\cn(u-v)
& = &
\dfrac{\cn^2 u - \dn^2 u\,\sn^2 v}{1-k^2\,\sn^2 u\,\sn^2 v}
\;,
\vphantom{\Bigg|}\cr
\dn(u+v)\,\dn(u-v)
& = &
\dfrac{\dn^2 u - k^2\,\cn^2 u\,\sn^2 v}{1-k^2\,\sn^2 u\,\sn^2 v}
\;.
\vphantom{\Bigg|}
\label{AddtionSubtraction}
\end{eqnarray}
Therefore,
\begin{eqnarray}
\sn(u+v)\,\sn(u-v) + \cn(u+v)\,\cn(u-v)
& = & 
1 -\dfrac{2\dn^2 u\,\sn^2 v}{1-k^2\,\sn^2 u\,\sn^2 v}
\;,
\vphantom{\Bigg|}\cr
k^2\sn(u+v)\,\sn(u-v) + \dn(u+v)\,\dn(u-v)
& = &
1 -\dfrac{2k^2\cn^2 u\,\sn^2 v}{1-k^2\,\sn^2 u\,\sn^2 v}
\;.
\vphantom{\Bigg|}
\label{AdditionSubtraction2}
\end{eqnarray}
When $v=0$, these reduce to Eq.~\eqref{PythagoreanAnalog}.
Similarly, we find
\begin{eqnarray}
\sn(u+v)\cn(u-v)-\cn(u+v)\sn(u-v)
& = &
\dfrac{2\dn u\,\cn v\,\sn v}
{1-k^2\,\sn^2 u\,\sn^2 v}
\;,\vphantom{\Bigg|}\cr
\sn(u+v)\dn(u-v)-\dn(u+v)\sn(u-v)
& = &
\dfrac{2\cn u\,\dn v\,\sn v}
{1-k^2\,\sn^2 u\,\sn^2 v}
\;,\vphantom{\Bigg|}\cr
\cn(u+v)\dn(u-v)-\dn(u+v)\cn(u-v)
& = &
-\dfrac{2(1-k^2)\sn u\,\sn v}
{1-k^2\,\sn^2 u\,\sn^2 v}
\;,\vphantom{\Bigg|}
\label{AdditionSubtraction3}
\end{eqnarray}

\item Lambert series expansion:
Let
\begin{equation}
q\;=\;\exp[-\pi K'/K]\;,
\end{equation}
where
\begin{eqnarray}
K\;=\; K(k^2) & = & \int_0^1\dfrac{dx}{\sqrt{(1-x^2)(1-k^2 x^2)}}\;,\cr
K'\;=\; K'(k^2) \;=\; K(k^{\prime 2}) & = & \int_0^1\dfrac{dx}{\sqrt{(1-x^2)(1-k^{\prime 2} x^2)}}\;,
\end{eqnarray}
and $k'=\sqrt{1-k^2}$. 
We have
\begin{eqnarray}
\sn(u,k) & = & \dfrac{2\pi}{K}\dfrac{1}{k}\sum_{n=0}^{\infty}
\dfrac{q^{n+1/2}}{1-q^{2n+1}}\;\sin((2n+1)v)
\;,\cr
\cn(u,k) & = & \dfrac{2\pi}{K}\dfrac{1}{k}\sum_{n=0}^{\infty}
\dfrac{q^{n+1/2}}{1+q^{2n+1}}\;\cos((2n+1)v)
\;,\cr
\dn(u,k) & = & \dfrac{2\pi}{K}\left[
\dfrac{1}{4} + \sum_{n=0}^{\infty}
\dfrac{q^{n}}{1+q^{2n}}\;\cos(2nv)
\right]
\;,
\end{eqnarray}
where
\begin{equation}
v \;=\; \dfrac{\pi}{2K}\,u\;.
\end{equation}
$2\pi/K$ and $q$ can also be expanded as
\begin{eqnarray}
\dfrac{2\pi}{K}
& = &
4 - k^2 - \dfrac{5k^4}{16} - \dfrac{11 k^6}{64}
+ \cdots
\;,
\vphantom{\Bigg|}
\cr
q & = &
\dfrac{k^2}{16} 
+ \dfrac{k^4}{32}
+ \dfrac{21 k^6}{1024}
+ \dfrac{31 k^8}{2048}
+ \cdots 
\vphantom{\Bigg|}
\end{eqnarray}
which leads to
\begin{eqnarray}
\sn(u,k)
& = & 
\bigg(
1+\dfrac{k^2}{16}+\dfrac{7k^4}{256}+\cdots
\bigg)\sin v
+\bigg(
\dfrac{k^2}{16}+\dfrac{k^4}{32}+\cdots
\bigg)\sin(3v)
+\bigg(
\dfrac{k^4}{256}+\cdots
\bigg)\sin(5v)
+ \cdots
\vphantom{\Bigg|}\cr
\cn(u,k)
& = &
\bigg(
1-\dfrac{k^2}{16}-\dfrac{9k^4}{256}-\cdots
\bigg)\cos v
+\bigg(
\dfrac{k^2}{16}+\dfrac{k^4}{32}+\cdots
\bigg)\cos(3v)
+\bigg(
\dfrac{k^4}{256}+\cdots
\bigg)\cos(5v)
+ \cdots
\vphantom{\Bigg|}\cr
\dn(u,k)
& = &
\bigg(
1 - \dfrac{k^2}{4} - \dfrac{5k^4}{64} - \cdots
\bigg)
+\bigg(
\dfrac{k^2}{4} + \dfrac{k^4}{16} + \cdots
\bigg)\cos(2v)
+\bigg(
\dfrac{k^4}{64} + \cdots
\bigg)\cos(4v)
+ \cdots
\vphantom{\Bigg|}
\end{eqnarray}

\end{itemize}

\bibliographystyle{apsrev4-2}
\bibliography{Nambu}

\end{document}